\documentclass[fleqn,usenatbib]{mnras}

\usepackage{newtxtext,newtxmath}

\usepackage[T1]{fontenc}

\DeclareRobustCommand{\VAN}[3]{#2}
\let\VANthebibliography\thebibliography
\def\thebibliography{\DeclareRobustCommand{\VAN}[3]{##3}\VANthebibliography}

\newcommand{\labsim}[1]{\fontfamily{qcr}\selectfont \textbf{#1}}

\usepackage{graphicx}	
\usepackage{amsmath}	
\usepackage{cleveref}
\usepackage{float}

\title[Spin-dependent MBH pair feedback]{Dynamical evolution of massive
black hole pairs in the presence of spin-dependent radiative feedback}

\author[F. Bollati et al.]{
Francesco Bollati,$^{1,2}$\thanks{E-mail: fbollati@uninsubria.it}
Alessandro Lupi,$^{3,2}$
Massimo Dotti,$^{3,2,4}$
Francesco Haardt$^{1,2,4}$
\\
$^{1}$DiSAT, Universit\`a degli Studi dell'Insubria, via Valleggio 11, I-22100 Como, Italy\\
$^{2}$INFN, Sezione di Milano-Bicocca, Piazza della Scienza 3, I-20126 Milano, Italy\\
$^{3}$Dipartimento di Fisica G. Occhialini, Universit\`a di Milano-Bicocca, Piazza della Scienza 3, I-20126 Milano, Italy\\
$^{4}$INAF, Osservatorio Astronomico di Brera, Via E. Bianchi 46, I-23807 Merate, Italy
}
\date{Accepted XXX. Received YYY; in original form ZZZ}

\defcitealias{Cenci2021}{C21}
\defcitealias{Sala2021}{S21}

\pubyear{2015}

\begin{document}
\label{firstpage}
\pagerange{\pageref{firstpage}--\pageref{lastpage}}
\maketitle

\begin{abstract}
\noindent
The putative ubiquity of massive black holes (MBH) at the center of galaxies, and the hierarchical progress of structure formation along the cosmic history, together necessarily imply the existence of a large population of cosmic MBH binaries. Such systems are understood to be the loudest sources of gravitational waves at mHz frequencies, the regime that will be probed by the next Laser Interferometer Space Antenna (LISA). It has been proposed that the rate at which MBHs pair and then bind to form binaries is critically dependent upon the feedback exerted by the MBHs on the surrounding gaseous environment. Using the publicly available code \textsc{gizmo}, we perform a suite of simulations aimed at studying the dynamics of a MBH pair embedded in a gaseous disk on $\sim$ 100 pc scale. By means of dedicated modules, we follow the dynamics of MBHs in the presence of diﬀerent spin-dependent radiative feedback models, and compare the results to a benchmark case with no feedback at all. Our main ﬁnding is that feedback causes the secondary MBH to shrink its orbit at a reduced pace, when compared to models where feedback is absent. Moreover, such slower inspiral occurs on eccentric orbits, as feedback has the net eﬀect of hampering the circularization process. Though idealized in many aspects, our study highlights and quantiﬁes the importance of including spin-dependent feedback recipes in hydrodynamic simulations of MBH pairs, and ultimately in assessing the cosmological coalescence rate of such systems in view of their detection through gravitational waves.

\end{abstract}




\section{Introduction}

Massive black hole (MBH) pairs are expected to form during galaxy mergers (\citealt{BBT80}, see \citealt{derosa} for a recent review). During such dramatic interactions large amounts of gas are driven toward the centre of the forming remnant, either due to gravitational  \citep[e.g.][]{hernquist89,bh91,bh96,mh96} or hydrodynamical \citep[][]{barnes02,CD17,BB18} torques. Such inflows result in the formation of a massive circum-nuclear (gaseous) disc (CND) commonly observed in the late-stages of a galaxy merger \citep[e.g.][]{SM96,DS98}.

Early numerical studies suggested that dynamical friction (DF) exerted by the dense CND onto the pairing MBHs may act in speeding-up their orbital decay \citep[][]{escala05,dotti06}, at the same time circularizing the decaying orbits \citep[][]{Dotti2007}. As a consequence, the delays between the galaxy merger completion and the formation of the MBH binary (MBHB) would be negligibly small even at high ($z\gtrsim 6$) redshifts, impacting the redshift distribution of the expected MBHB coalescences detectable by the future Laser Interferometer Space Antenna \citep[LISA,][]{lisa}. Furthermore, the drag toward circular corotation due to the CND-driven DF ensures small relative velocities between the pairing MBHs and the surrounding gas. This very fact enhances the probability of detecting dual AGNs on small scales \citep[$\lesssim 1$ kpc][]{dotti09}, and promotes the prompt alignment of MBH spins with the orbital angular momentum \citep[][]{dotti10}, hence decreasing the expected recoil velocities at coalescence \citep[][]{bogdanovic07}. The net effect is an increase of the MBH occupation fraction at all redshifts \citep[][]{volonteri10}.

The above-mentioned studies, however, did not consider the possible feedback that accreting MBHs would exert onto the surrounding gas, altering its local properties and, therefore, its dynamical friction effect. Indeed, early indications of a sign reversal of the DF torque exerted by a rotating gaseous background affected by MBH feedback has been discussed on galactic scales by \cite{sijacki11}. The authors found that recoiling MBHs on initially radial orbits
tend to circularize corotating with the gaseous disc, and experience a fast orbital decay when AGN feedback is not included, while settle on counter-rotating orbits, resulting in higher relative velocities with respect to the gaseous environment and long orbital decay timescales, when accretion-powered feedback is included.  

This early claim has been numerically confirmed in the contexts of MBH pair dynamics on CND scales \citep[see, e.g.,][]{rafael17}, where the pairing efficiency of the MBH is significantly reduced by the AGN feedback and some small acceleration in the direction opposite to that of the standard DF is measured.\footnote{See, however, \citet{2020MNRAS.496.1909T} for an opposite indication when considering high densities ($\gtrsim 10^6$ cm$^{-3}$), large relative velocities ($\gtrsim 100$ km s$^{-1}$) between the black hole and the gas, and the effect of dust in determining the extent of the region affected by feedback.} Similar indications have been found for planetary migration in protoplanetary discs \citep[][]{masset17b,masset17} and have been discussed analytically in diverse contexts by \cite{2020MNRAS.492.2755G}.

In all the studies (both numerical or analytical) including AGN feedback mentioned above the energy or momentum injection has been assumed to be isotropic. However, at sufficiently small scales, the feedback may have significant deviations from isotropy, both in case of direct radiative feedback from the MBH accretion disc, as well as wind-mediated outflow or kinetic feedback from relativistic jets \citep[see, e.g., the discussion in][and references therein]{2020MNRAS.496.1909T}. 
An anisotropic feedback would increase significantly the complexity of the problem and, for some specific configurations, it may decrease the effect of feedback onto the local gaseous environment if the majority of the energy-momentum of the outflows escapes through a minimal-resistance path before altering the gas dynamics. 

Here, we consider for the first time the impact of a spin-dependent radiative feedback on the dynamics of a MBH pair embedded in a CND, exploring different feedback models based on an anisotropic momentum injection in the gas. In particular, we couple the direction of the feedback with that of each MBH spin, evolving due to gas accretion following the prescription discussed in \citealt[][]{Cenci2021} and \citealt[][]{Sala2021}, hereafter \citetalias{Cenci2021} and \citetalias{Sala2021}, respectively. In \S~\ref{sec:model}, we describe the model of spin-dependent feedback we adopt in our simulation suite. The specific set up of our simulations is presented in \S~\ref{sec:setup}, while results are discussed in \S~\ref{sec:results}. Finally, \S~\ref{sec:summary} is devoted to summary and concluding remarks.

\section{Model}\label{sec:model}

In this section we review the model for MBH spin evolution and feedback implemented in \textsc{gizmo} \citep{Hopkins2015} by  \citetalias{Cenci2021} and \citetalias{Sala2021}.

The MBH particle is meant to represent a structured, sub-resolution system consisting of a MBH surrounded by an unresolved, warped accretion $\alpha$-disc \citep{SS73}. The MBH particle is completely characterized by its dynamical mass $M_{\bullet, \textrm{dyn}}$, \footnote{The dynamical mass is that used in the computation of the gravitational force.} whereas the sub-resolution, proper MBH (hereafter, simply the MBH)  by its mass $M_\bullet$ and dimensionless spin-parameter $a = c|\mathbf{J}_\bullet |/ G M_\bullet ^2$, where $\mathbf{J}_\bullet$ is the MBH angular momentum, $c$ the speed of light and $G$ the gravitational constant. The unresolved accretion disc is specified by its mass $M_\alpha$, its total angular momentum $\mathbf{J}_\alpha$, and
the accretion rate $\dot{M}_\textrm{acc} = f_\textrm{Edd} \dot{M}_\textrm{Edd}$, where $\dot{M}_\textrm{Edd} = 4\pi G M_\bullet m_\textrm{p} / (\sigma_\textrm{T}\eta c)$ is the Eddington accretion rate, $m_\textrm{p}$ the proton mass,  $\sigma _\textrm{T}$ the Thomson scattering cross-section and $\eta$ the disc radiative efficiency. 
In general, $\mathbf{J}_\bullet$ and $\mathbf{J}_\alpha$ are misaligned, i.e., the $\alpha$-disc is warped, with the inner region laying in the MBH equatorial plane and the outer part aligned with $\mathbf{J}_\alpha$ \citep{BP75}. 
The $\alpha$-disc model employs prescriptions for the radial and vertical viscosity $\nu _1$ and $\nu _2$, which regulate respectively the accretion onto the MBH and the propagation of vertical perturbations. Both viscosities are expressed in terms of the Shakura \& Sunyaev $\alpha$ parameter \citep{SS73,LP2006}.

BH and unresolved disc parameters are updated every MBH timestep
according to analytical prescriptions that link the sub-resolution system to the resolved scales. The unresolved system parameters, in turn, are used to model the effects of MBH feedback on resolved scales. 
The coupling between resolved and unresolved scales is limited to those particles lying within the MBH smoothing kernel, which is defined as a spherical region centered on the MBH enclosing a given effective number of particles $N_{\textrm{ngb},\bullet}$. In order to avoid coupling feedback on very large scales, the kernel size is capped at a maximum radius $R_{\bullet,\textrm{max}}$. The above-mentioned prescriptions have been implemented in \textsc{gizmo} by \citetalias{Cenci2021} and \citetalias{Sala2021}, and their main features are summarized in the following.

\subsection{Sub-grid accretion and spin evolution}\label{cASE}

The time evolution of the MBH mass is governed by the accretion rate $\dot{M}_\textrm{acc}$ and the accretion radiative efficiency $\eta$, whereas the mass of the unresolved disc feeding the MBH evolves according to the mass inflow $\dot{M}_\textrm{in}$ from resolved scales, the mass outflow $\dot{M}_\textrm{w}$, and  $\dot{M}_\textrm{acc}$ as  
\begin{align}
 & \dot{M}_\bullet = (1-\eta)\dot{M}_\textrm{acc}, \label{MBHdot} \\
 & \dot{M}_\alpha = \dot{M}_\textrm{in} - \dot{M}_\textrm{acc} - \dot{M}_\textrm{w}. \label{Malphadot}
\end{align}
Equations (\ref{MBHdot}) and (\ref{Malphadot}) are then used to update the masses of MBH and disc after a time-step $\Delta t$ as $M_{\bullet,t+\Delta t} = M_{\bullet,t} + \dot{M}_{\bullet,t} \Delta t$ and 
$M_{\alpha,t+\Delta t} = M_{\alpha,t} + \dot{M}_{\alpha,t} \Delta t$.
$\dot{M}_\textrm{in}$, the mass inflow onto the MBH particle, is modelled as spherical accretion a la Bondi-Holyle-Littleton
\footnote{We point out that the classical Bondi-Hoyle accretion we employ (Eq. \ref{Bondi}) has the tendency to overestimate the accretion on the BH, as shown by \cite{Hopkins11}, \cite{Curtis16} and \cite{Tremmel17}.
} \citep{HL39, Bondi44,Bondi52}, implemented by \cite{Springeletal2005} as
\begin{equation}
    \dot{M}_\textrm{in} = \frac{4\pi G^2 M_\bullet^2\rho}{(c_\textrm{s}^2+|\mathbf{v}_\bullet - \mathbf{v}|^2)^{3/2}},
    \label{Bondi}
\end{equation}
where $\rho$, $\mathbf{v}$ and $c_\textrm{s}$ are the gas density, velocity and sound speed, respectively, computed as mass-weighted averages on the gas particles within the MBH smoothing kernel, and $\mathbf{v}_\bullet$ is the MBH velocity.
In \citetalias{Cenci2021}'s implementation, the disc mass $M_\alpha$ is allowed to vary between a user-defined minimum value $M_{\alpha, \rm seed}$, which is reset in case $M_\alpha$ vanished due to accretion onto the MBH, and a maximum value set to prevent the unresolved disc from becoming self-gravitating.

$\dot{M}_\textrm{acc}$ is self-consistently evolved according to the evolution of the sub-grid quantities  $ \{ \mathbf{J}_\bullet, \mathbf{J}_\alpha, \eta, f_\textrm{Edd} \}$, which are updated at  each MBH time-step. The time variation of the MBH angular momentum, $\dot{\mathbf{J}}_\bullet$, is determined by the angular momentum carried by the unresolved accreted gas at the innermost stable orbit (ISCO), which modifies the spin magnitude, and by the gravitomagnetic torque between the MBH spin and disc angular momentum, which tends to align the MBH spin to the total (i.e., MBH + disc) angular momentum \citep{king2005, Fiacconi2018}. Conservation of angular momentum implies $\dot{\mathbf{J}}_\alpha = -\dot{\mathbf{J}}_\bullet + \dot{\mathbf{J}}_\textrm{in}$,
where $\dot{\mathbf{J}}_\textrm{in}$ is the angular momentum inflow from the resolved gas, i.e., 
$\dot{\mathbf{J}}_\textrm{in}=  \dot{M}_\textrm{in}\mathbf{\Lambda}_\textrm{in}$, where $\mathbf{\Lambda}_\textrm{in}$ is the angular momentum per unit mass of the inflowing material \citepalias{Cenci2021}. Then, the MBH and disc angular momenta are updated as $\mathbf{J}_{\bullet, t + \Delta t} = \mathbf{J}_{\bullet, t} + \dot{\mathbf{J}}_{\bullet,t} \Delta t$ and  
    $\mathbf{J}_{\alpha, t + \Delta t} = \mathbf{J}_{\alpha, t} + \dot{\mathbf{J}}_{\alpha,t} \Delta t$.
The radiative efficiency $\eta$ depends on the location of the ISCO which, in turn, is a function of the MBH spin parameter $a$, and is then consistently evolved. Finally, once the subgrid parameters are updated, $f_\textrm{Edd}$ can be computed following the prescription by \cite[][see their Eq. 2]{Fiacconi2018}, hence giving $\dot{M}_\textrm{acc}$.
In the scheme just described, the MBH timestep $\Delta t$ is taken small enough to resolve the sub-grid accretion and spin evolution and large enough to guarantee that the disc attains a steady-state warped profile, as assumed in our prescriptions. 

The outflow rate $\dot{M}_\textrm{w}$ is instead computed from the unresolved system parameters as
\begin{equation}
    \dot{M}_\textrm{w} v_\textrm{w} = p L_\textrm{bol} / c = p \eta c \dot{M}_\textrm{acc},
    \label{wind}
\end{equation}
where $L_\textrm{bol}$ = $\eta \dot{M}_\textrm{acc}c^2$ is the disc bolometric luminosity, $v_\textrm{w}$ the wind speed, and $p$ is the ratio between the wind and disc radiation momentum fluxes. Both $v_\textrm{w}$ and $p$ are free parameters of the model.

 \subsection{Stochastic feedback}\label{cSF}

The last quantity we need to evolve is the dynamical mass of the MBH particle. While $M_\bullet$ and $M_\alpha$ evolve smoothly over time, to keep under control the error made in mass conservation, $M_{\bullet,\textrm{dyn}}$ is instead subject to a stochastic evolution \citep{Springeletal2005}. This is modeled via a stochastic selection of gas particles within the MBH kernel, whose mass is reduced by a fraction $f$, which is added to the MBH dynamical mass. 
The remaining $(1-f)$ fraction of the selected particles is then kicked outwards with a velocity $v_\textrm{w}$ along specific directions that depend on the chosen feedback model, thus concurring to form the resolved MBH-driven wind. The fraction $f$ is defined as $f = 1-\dot{M}_\textrm{w}\Delta t/(\sum _k^N m_k)$, where $m_k$ is the mass of the k-th gas particle among the N selected. This choice of $f$ guarantees that the entire amount of ejected mass $(1-f)\sum^N m_k$ is, at every time-step, equal to $\dot{M}_\textrm{w}\Delta t$.
The probabilities to select particles are chosen to guarantee that, on average, the mass transferred to $M_{\bullet,\textrm{dyn}}$ is $f\sum _k^N m_k \simeq M_{\bullet,t+\Delta t}+M_{\alpha,t+\Delta t} - M_{\bullet,t} - M_{\alpha,t}$ (see \citetalias{Sala2021} for further details).\footnote{In order to ensure that the dynamical mass follows on average the physical mass, when $M_{\bullet,t+\Delta t} < M_{\bullet,\textrm{dyn}}$ we only change the momentum of the selected particles, and leave the MBH dynamical mass unchanged.}

By varying the kick direction, the outflow anisotropy can be tuned to reproduce different feedback mechanisms. In particular, the outflow can be either modelled as an isotropic wind, where the selected particles receive a kick along the radial direction, as a collimated jet parallel to the gas angular momentum, as implemented by \cite{DAA2017}, or it can be assumed to have a biconical shape, as implemented by \citetalias{Sala2021}. In the latter case, the kick direction is randomly sampled within a cone of given semi-aperture $\theta _\textrm{bic}$, with the cone axis either fixed in time or consistently evolved during the simulation (e.g., parallel to the MBH spin).


\section{Numerical simulation setup}
\label{sec:setup}

 In order to simulate the dynamics of MBH pairs in presence of spin-dependent radiative feedback, in this work we employed the publicly available N-body, mesh-less hydrodynamic code \textsc{gizmo} \citep{Hopkins2015} supplied with the implementations by \citetalias{Cenci2021} and \citetalias{Sala2021}, presented in \Cref{cASE,cSF}. This enabled us to investigate the role of feedback in the orbital evolution of MBH pairs placed in a gaseous and stellar environment. Simulations were run on the CINECA cluster MARCONI 100.


Here we discuss the setup of the numerical simulations we performed, consisting in a MBH pair embedded in a gaseous CND and in a stellar bulge. The initial conditions have been created by first initializing
the stellar and gaseous components in dynamical equilibrium with the primary MBH (placed at the center of the system) and by subsequently adding the secondary MBH to the relaxed system. We achieved the first step by using the publicly available code GD\_BASIC \citep{Lupi2015}, building up a `Bulge+CND+Primary' (BCP hereinafter) system characterized by:
\begin{itemize}
    \item a spherical stellar bulge described by an \cite{H90} profile
    \begin{equation}
        \rho _\textrm{b} (r) = \frac{M_\star}{2\pi}\frac{r_\star}{r(r+r_\star )^3},
        \label{Hernquist}
    \end{equation}
    where $r$ is the spherical radial coordiante, $M_\star = 5\times10^8 M_\odot$ the total bulge mass and $r_\star = 100$ pc the bulge scale radius;
    \item a rotationally supported exponential disc in vertical hydrostatic equilibrium  whose surface density profile is
    \begin{equation}
        \Sigma (R) = \frac{M_\textrm{d}}{2\pi R_\textrm{d}^2}e^{-R/R_\textrm{d}},
        \label{CND}
    \end{equation}
    where $R$ is the cylindrical radial coordinate, $R_\textrm{d} = 50$ pc the disc scale radius and $M_\textrm{d} = 10^8 M_\odot$ the disc total mass; 
    \item a primary MBH with dynamical mass $M_1 = 10^7 M_\odot$ at rest in the center of the system. 
\end{itemize}

The stellar and gaseous components are sampled  by  $N_\star = 5\times 10^6$ and $N_\textrm{d} = 10^6$ particles respectively, corresponding to a mass resolution of $10^2 M_\odot$ for both. The spatial resolution is determined by the Plummer equivalent gravitational softening $\epsilon_\textrm{soft}$. For stellar and MBH particles it is fixed at 0.1 pc and 0.33 pc, respectively, while for gas particles it is adaptively set equal to the hydrodynamic kernel size, i.e., the radius encompassing an effective number of neighbours $N_\textrm{ngb} = 32$, down to a minimum allowed value $\epsilon_\textrm{soft, min} = 0.1$.  The gas particles are also initialized with a uniform temperature $T=2\times 10^4$ K, assuming an ideal equation of state with adiabatic index $\gamma = 5/3$. Once created, in order to relax the system, the BCP is evolved for 20 Myr, corresponding to $\sim 6$ orbits at $R_\textrm{d}$ and $\sim 3.22$ orbits at $r_\star$.

After relaxation, we introduced a secondary MBH with dynamical mass $M_2$ in the disc plane ($z=0$) at a separation of 80 pc from the BCP center of mass\footnote{After adding the secondary MBH, we also shifted the positions and velocities of all the particles in order to move the center of mass of the BCP+secondary system at rest in the origin.}, producing different initial conditions depending on the initial mass ratio $q = M_2/M_1$ and initial eccentricity $e$.
These initial conditions are aimed at modeling the final stages of the MBHs DF-driven inspiral that brings the MBHs separation from kpc to pc scales \citep{Mayer07, LISAW}.
In our fiducial simulations (indicated with {\labsim{f}}) we initialized the secondary with $q = 1/2$ and $e=0$ (w.r.t. the center of mass of the BCP) and initial velocity $\mathbf{v}_2 (t=0) = \sqrt{R|d\Phi /d R|}\, \hat{\varphi}$, where $\hat{\varphi}$ is the azimuthal unit vector and $\Phi$ is the gravitational potential of the BCP. Compered to this fiducial runs, simulations labelled as {\labsim{q}} all have a lower mass ratio ($q=1/6$), simulations labelled as {\labsim{e}} have a non-vanishing initial radial velocity component that sets the initial eccentricity to $e=0.5$ (see Table \ref{Tab}). For each of these three initial set-ups ({\labsim{f, q}} and {\labsim{e}}) we performed four simulations considering different feedback models: i) the case without feedback (labelled as {\labsim{nofb}}), ii) isotropic feedback (labelled as {\labsim{iso}}), iii) biconical feedback with the cone axis fixed and parallel to the vertical direction $\hat{z}$ (labelled as {\labsim{z}}) and iv) biconical feedback with the cone axis aligned to the evolving MBH spin direction (labelled as {\labsim{a}}). In the latter two cases we fixed $\theta _\textrm{bic} = 45^\circ$ and, whenever feedback is present, we used $v_\textrm{w} = 500$ km/s and $p = 1$, i.e., the radiation momentum flux is entirely transferred to the wind. Our complete simulation suite therefore comprises a total of 12 runs. 

\begin{table}
\centering
    \caption{Summary of the parameters adopted in our simulations. \emph{Top:} parameters that vary across our simulation suite.  The following parameters are the same for all runs: for feedback launching $\theta_\textrm{bic} = 45^\circ$, $p=1$, $v_\textrm{w} = 500$ km/s and for the sub-grid system $a=0.5$, $f_\textrm{Edd} = 0.01$, $M_\alpha /M_\bullet = 0.005$ and $\alpha = 0.1$. Our choice of $M_\alpha$ guarantees that the initial disc mass is smaller than the disc self-gravitating mass $M_\textrm{sg}$.
    \emph{Bottom:} gravitational softening for the different components.}
    \begin{tabular}{ c | c c c c  c}
            & $q$ & $e$ &  $\frac{J_\alpha}{J_\bullet}$ & $\frac{M_\alpha}{M_\textrm{sg}}$ & $R_{\bullet,\textrm{max}}$ [pc]  \\
            \hline 
            {\labsim{f}} & 1/2 &  0   & 2.85 & 0.52 & 3 \\
            {\labsim{q}} & 1/6 & 0   & 4.84 & 0.40 & 1\\ 
            {\labsim{e}} & 1/2 & 0.5 &  2.85 & 0.52 & 3\\
    \end{tabular}
\centering
    \begin{tabular}{ l | c }
            type & $\epsilon_\textrm{soft}$ [pc]  \\
            \hline 
            gas & 0.1 \\
            bulge & 0.1  \\ 
            MBH & 0.33  \\
    \end{tabular}
    \label{Tab}       

\end{table}    
We remark that the modules for sub-grid accretion plus spin-evolution and stochastic feedback (if present) are switched on for the secondary MBH only. This means we are considering the impact of feedback from the secondary on its own dynamics without accounting for the possible effects the feedback from the primary MBH may have on the secondary one. This is justified by the fact that we are in a regime where the relative separations of the two MBHs is large compared to the local regions possibly affected by feedback.


In all simulations, we initialise the secondary MBH mass as $M_\bullet = M_2/1.005$, $M_\alpha = 0.005 M_\bullet$, such that $M_\bullet + M_\alpha = M_2 (\equiv M_{\bullet,\textrm{dyn}})$. The disc angular momentum direction is along the $z-$axis, while the initial MBH spin is ``flipped downwards'' at an angle $5\pi/6$, with magnitude $a=0.5$. The Eddington ratio is set at $f_\textrm{Edd} = 0.01$, which together with the other sub-grid parameters constrains the value of $J_\alpha /J_\bullet$ (Eq. 5 in \citetalias{Cenci2021}).
Finally, we use $N_{\bullet, \textrm{ngb}} = 3 N_\textrm{ngb}$.

\section{Results}
\label{sec:results}

We discuss now the outcome of the simulations we carried out, starting from {\labsim{f}}-simulations in \cref{fsim}.  In the subsequent sections we perform the same analysis for {\labsim{e}}- and {\labsim{q}}-simulations.

\subsection{{\labsim{f}}-simulations}\label{fsim}
\subsubsection{Qualitative analysis}
\begin{figure}
     \centering
     \includegraphics[scale=0.55]{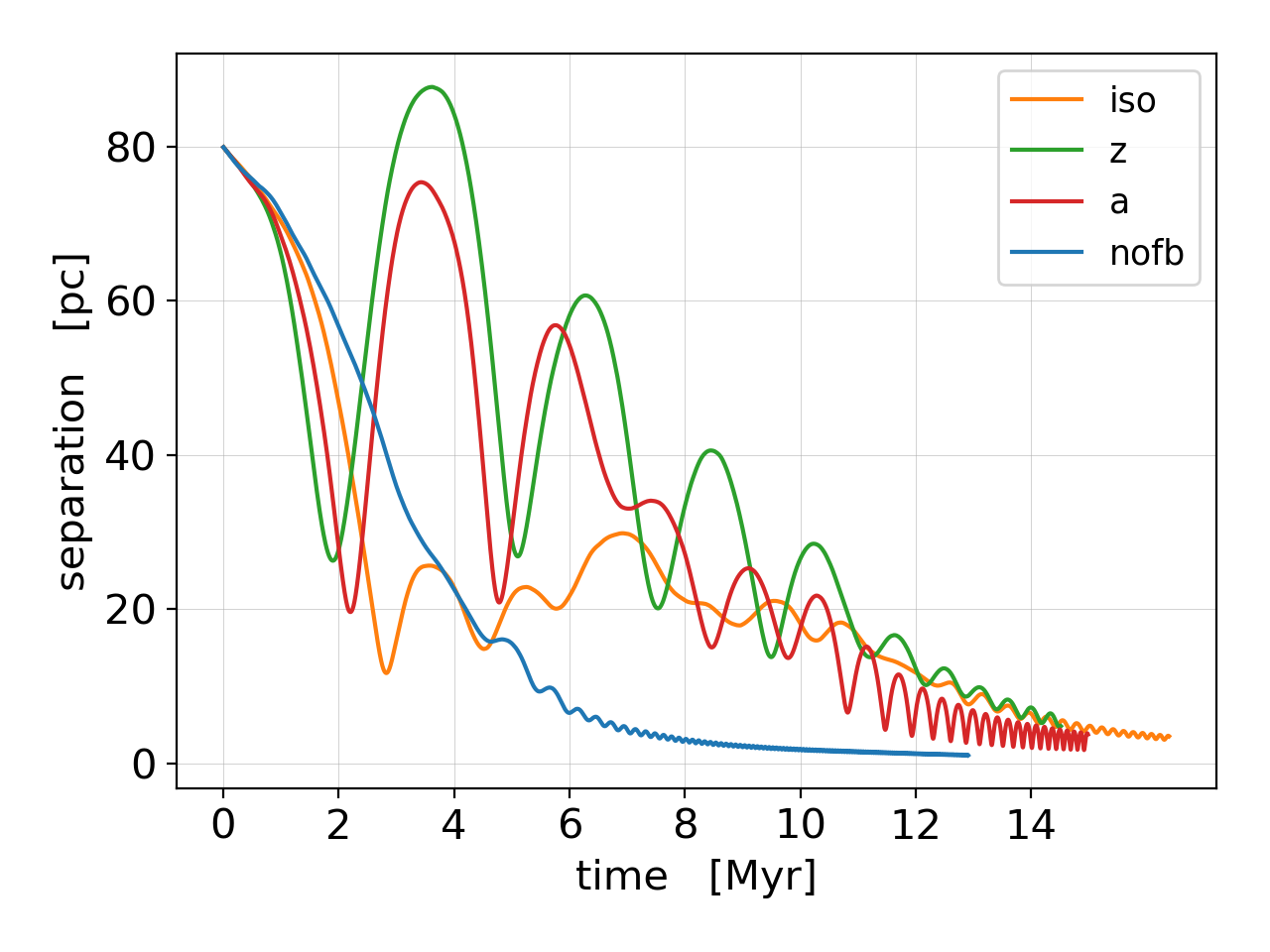}
     \caption{Time evolution of the MBH separation in {\labsim{f}}-simulations.}
     \label{distf}
 \end{figure}

\begin{figure}
     \centering
     \includegraphics[scale=0.55]{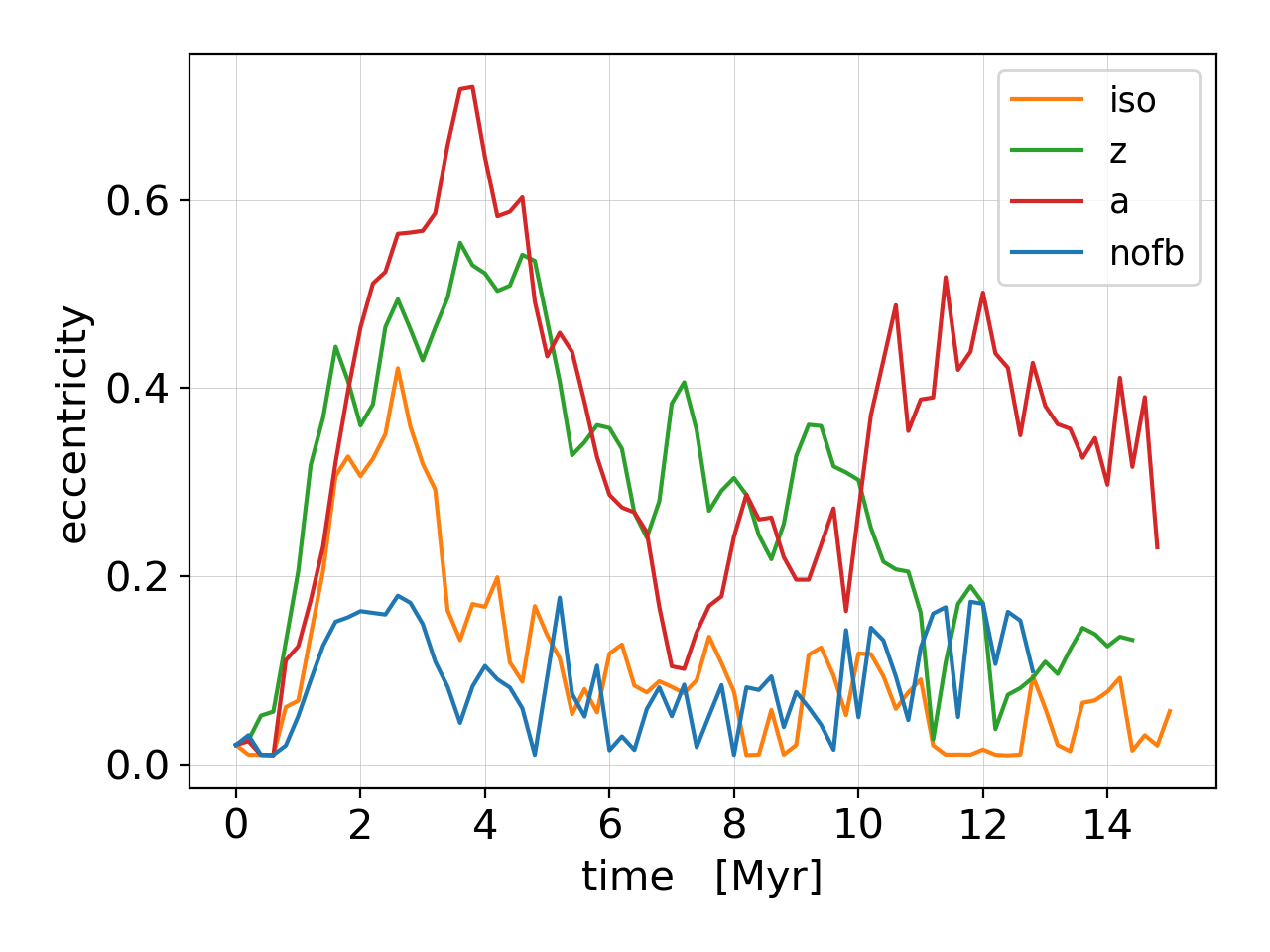}
     \caption{Time evolution of the eccentricity of $M_2$ in {\labsim{f}}-simulations.}
     \label{eccf}
 \end{figure}

\begin{figure}
     \hspace{-0.02\textwidth}
     \includegraphics[scale=0.55]{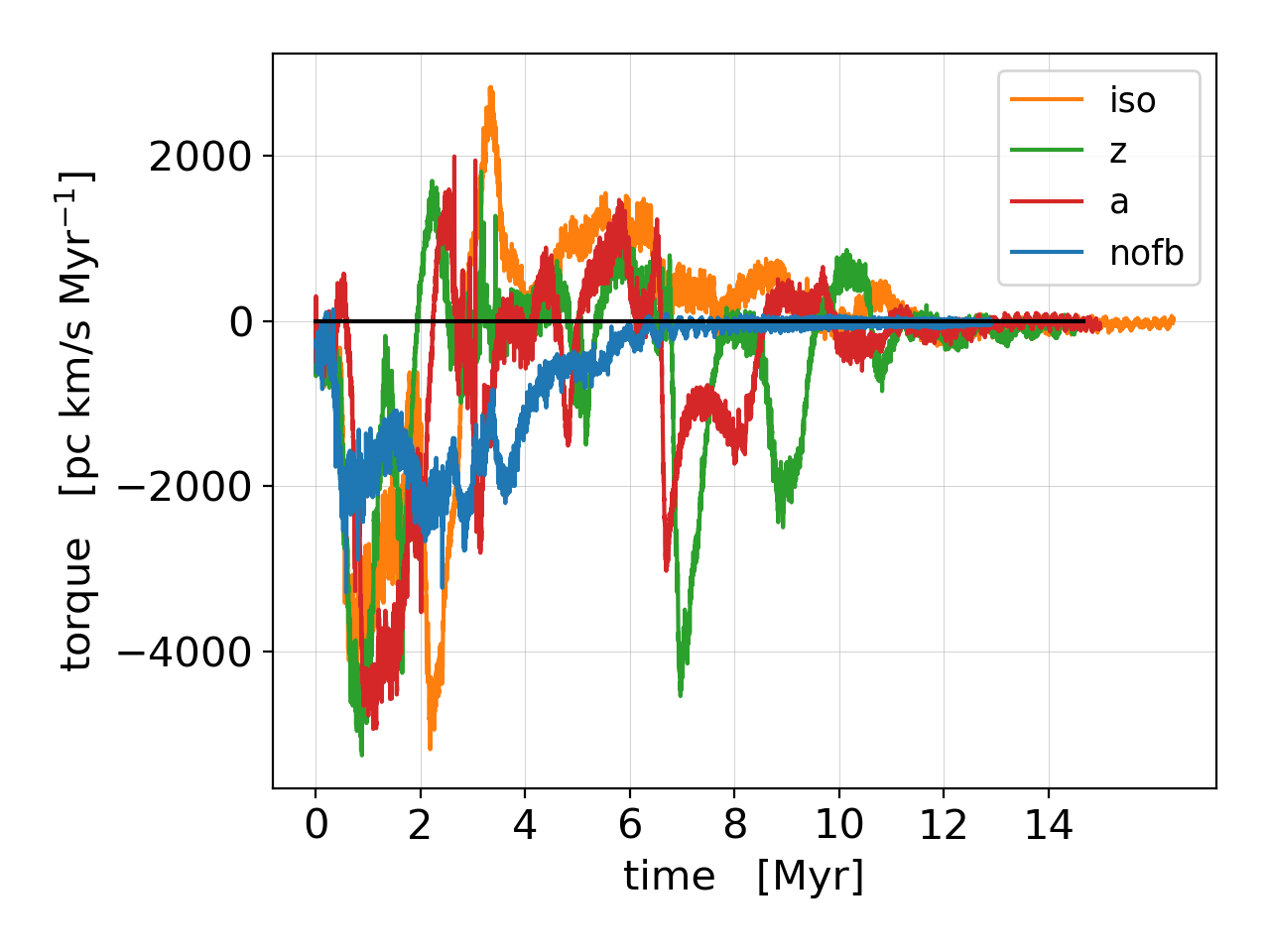}
     \caption{Time evolution of the $z$-component of the total torque acting on $M_2$ in {\labsim{f}}-simulations.}
     \label{torquesf}
 \end{figure}

The time evolution of the MBH separation and $M_2$ eccentricity is shown in  
Fig.~\ref{distf} and Fig.~\ref{eccf}, respectively, for the four tested different feedback models in {\labsim{f}}-simulations. Overall, we observe that in presence of feedback the timescale of orbital decay of $M_2$ is larger and the orbits tend to develop higher eccentricities. We can get some insights into such behavior from the evolution of the torques acting on $M_2$ shown in Fig.~\ref{torquesf}.
The torque in {\labsim{f\_nofb}} is always negative, indicating that efficient DF is acting on $M_2$, hence causing a net loss of angular momentum and energy,
leading to rapid inspiral towards $M_1$. On the other hand, switching feedback on, the torques on $M_2$ can become positive, indicating in these phases an inefficient (or even reversed) DF, i.e., a positive acceleration, and hence an orbital decay at slower pace. 

In more details, as  shown in Fig.~\ref{torquesf}, the $z$-component of the torque acting on $M_2$ initially quickly drops to negative values. In this phase, lasting $\sim 2$ Myrs, in {\labsim{f\_iso, f\_z}} and {\labsim{f\_a}} the torque is twice as large (in magnitude) than that in {\labsim{f\_nofb}}, i.e.,  DF is initially enhanced by feedback processes. In all cases, the initial larger loss of angular momentum in the presence of feedback is accompanied by an increase in eccentricity (Fig.~\ref{eccf}). Later on, the $z$-component of the torque becomes positive, i.e., the angular momentum increases, something not seen in {\labsim{f\_nofb}}, where the torque is always negative. 



We notice (see  Fig.~\ref{torquesf}) that in {\labsim{f\_iso}} the initial negative torque phase is prolonged slightly further compared to the other feedback simulations, causing the first pericenter to be closer to $M_1$ ($\sim 10$ pc), and to occur at a later time ($\sim 3$ Myr). Because of  such longer journey to the pericenter, $M_2$ looses more energy compared to the {\labsim{f\_z}} and {\labsim{f\_a}} cases, consequently acquiring a smaller eccentricity, $e \simeq 0.1$, with the separation stalled around $\simeq 20$ pc for the subsequent $\simeq 10$ Myr. In this phase, $M_2$ is subject to a net positive torque that traces a `reversed' DF. Then, the system enters a further phase in which the separation decreases again.

In {\labsim{f\_z}}, the eccentricity grows to $e\simeq 0.5$ when $M_2$ reaches the first apocenter, and then slowly decays. We also observe that the first apocenter is located at a distance larger than the initial MBHs separation, signaling a net gain of energy. In {\labsim{f\_a}}, $M_2$ orbits follow a similar trend, with an initial rapid rise in eccentricity followed by a slower circularization along $M_2$ orbital decay. 

Finally, we see that the two MBHs form a binary, at $\sim 5$ Myr in {\labsim{f\_nofb}}, and between 10-12 Myr in feedback simulations.
We comment the convergence of these results in the  \cref{Conv}.


\subsubsection{Quantitative analysis}\label{cA}

DF is generally attributed to the many two-body encounters between a massive object ($M_2$ in the present context) and background stars \citep{Chandra43}. In the case of a gaseous background, DF can be understood in terms of the gravitational interaction between $M_2$ and the density wake generated by its motion in the gaseous backround. In fact, the relative motion of $M_2$ with respect to the background creates an overdensity on the side opposite to the relative velocity between $M_2$ and fluid. The gaseous DF acts in all respect as a non-conservative drag force \citep{Ostriker99}. However, the gas around the MBH can be (partially) blown away by radiative feedback, thus hampering the formation of the density wake itself. The density enhancement trailing $M_2$ is, in this case, replaced by an underdensity, we refer to it as ``density bubble'', affecting $M_2$ dynamics in a decisive manner.

\begin{figure}
    \hspace{-0.03\textwidth}
    \includegraphics[scale=0.35]{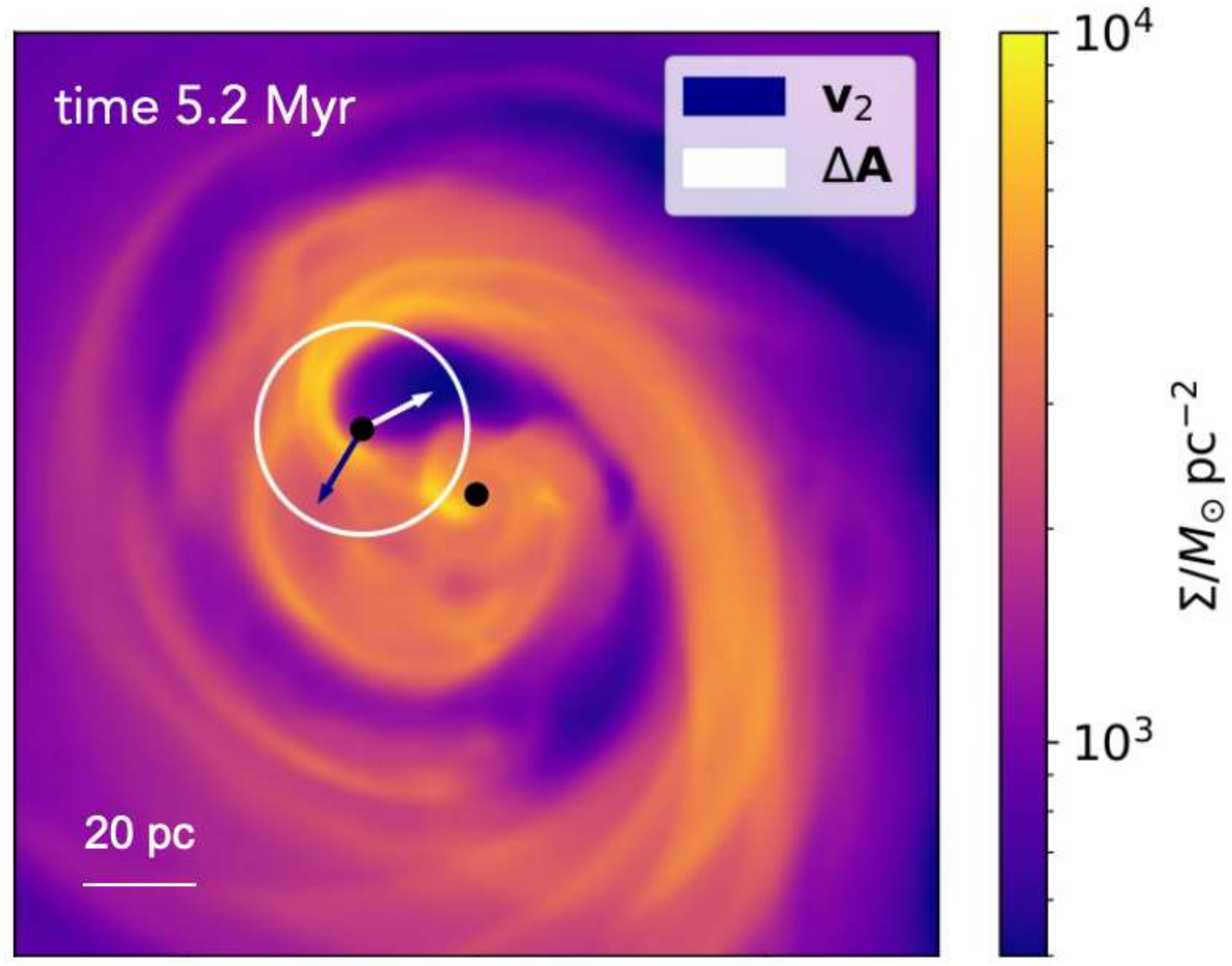}
    \caption{The surface density $\Sigma$ (in $M_\odot /{\rm pc}^2$) in a snapshot of  run {\labsim{f\_iso}}. The two black dots indicate the positions of the MBHs.
    The white circle bounds the region around $M_2$ used to define the anisotropy vector $\mathbf{A}$, indicated by the white arrow. The dark-blue arrow indicates $\mathbf{v}_2$, the $M_2$ velocity.}
    \label{A1fig}
\end{figure}

In order to quantify the effects the radiative feedback has on $M_2$ orbital decay, we introduce an ``anisotropy vector'' defined as  
\begin{equation}
   \mathbf{A} \equiv \sum_i m_i\, w(r_i) \frac{\mathbf{r}_i}{r_i}. 
\label{A11}
\end{equation}
Here, the sum is intended over all particles (with mass $m_i$ and position vector $\mathbf{r}_i$ in the CND  
plane centered in $M_2$) within a distance from $M_2$ equal to the minimum between 30 pc and the MBHs separation. Each particle is weighted by the force softening function implemented in \textsc{gizmo}, $w$ (see Appendix~\ref{appw}). As $\mathbf{A}$ evolves in time through coordinates $\mathbf{r}_i$, we consider the (normalised) difference $
    \Delta{\mathbf{A}}\equiv \mathbf{A}_0-\mathbf{A}$, 
where $\mathbf{A}_0$ is the anisotropy vector computed by considering, at each time, the current $M_2$ position but the initial distribution of gas. This allows us to quantify the time evolution of the anisotropy due to the MBH-gas interaction independently of any possible anisotropy already present at the beginning of the simulation.\footnote{We note that this choice would be slightly affected by the Poisson noise in the initial distribution of the gas.} The direction of $\Delta{\mathbf{A}}$ indicates the axis along which the anisotropy develops, pointing towards the lower density side. Therefore, in the presence of feedback, $\Delta{\mathbf{A}}$ indicates the bubble location, while, in absence of feedback, $\Delta{\mathbf{A}}$ points in the direction opposite to the over-density wake. Fig \ref{A1fig} shows $\Delta{\mathbf{A}}$ in a snapshot of {\labsim{f\_iso}}. 

If we now consider $A_2\equiv \Delta\mathbf{A} \cdot \mathbf{v}_2$, i.e., the projection of $\Delta\mathbf{A}$ along the $M_2$ velocity vector $\mathbf{v}_2$, we see that a positive value of $A_2$ would indicate that the bubble lies in front of $M_2$ (or that the overdensity lies behind $M_2$ in {\labsim{nofb}}-type simulations). In this case the gas distribution around $M_2$ exerts a gravitational force opposite to the direction of motion, resulting in an efficient DF. On the other hand, the underdense bubble lies behind $M_2$ for negative values of $A_2$, thus imparting a net acceleration to $M_2$ (see  Fig.~\ref{A1fig}).  

\begin{figure}
    \hspace{-0.03\textwidth}
    \includegraphics[scale=0.55]{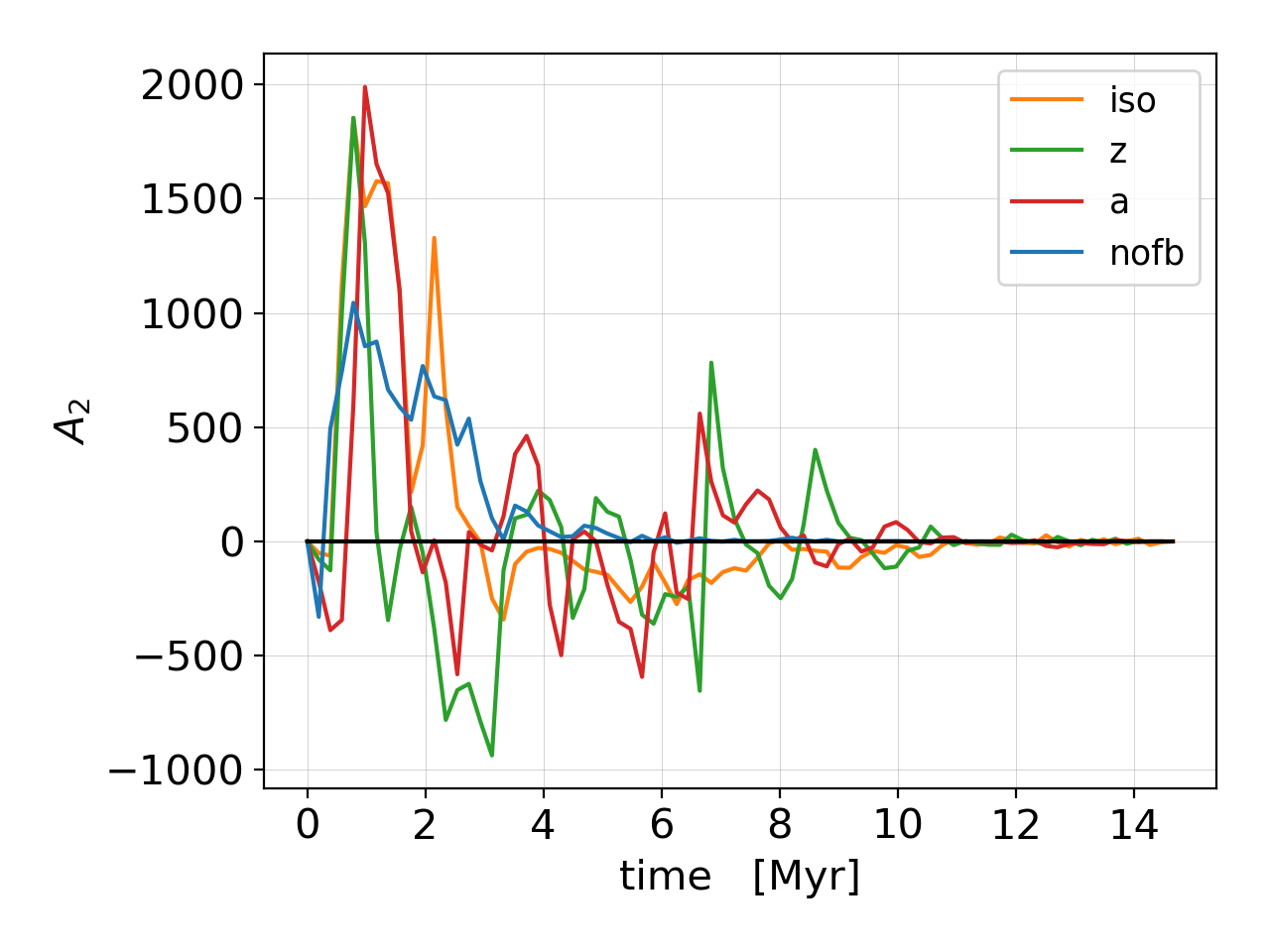}
    \caption{Time evolution of the anisotropy projection $A_2$ in {\labsim{f}}-simulations.}
    \label{Asf}
\end{figure}
\begin{figure}
    \hspace{-0.06\textwidth}
    \includegraphics[scale=0.38]{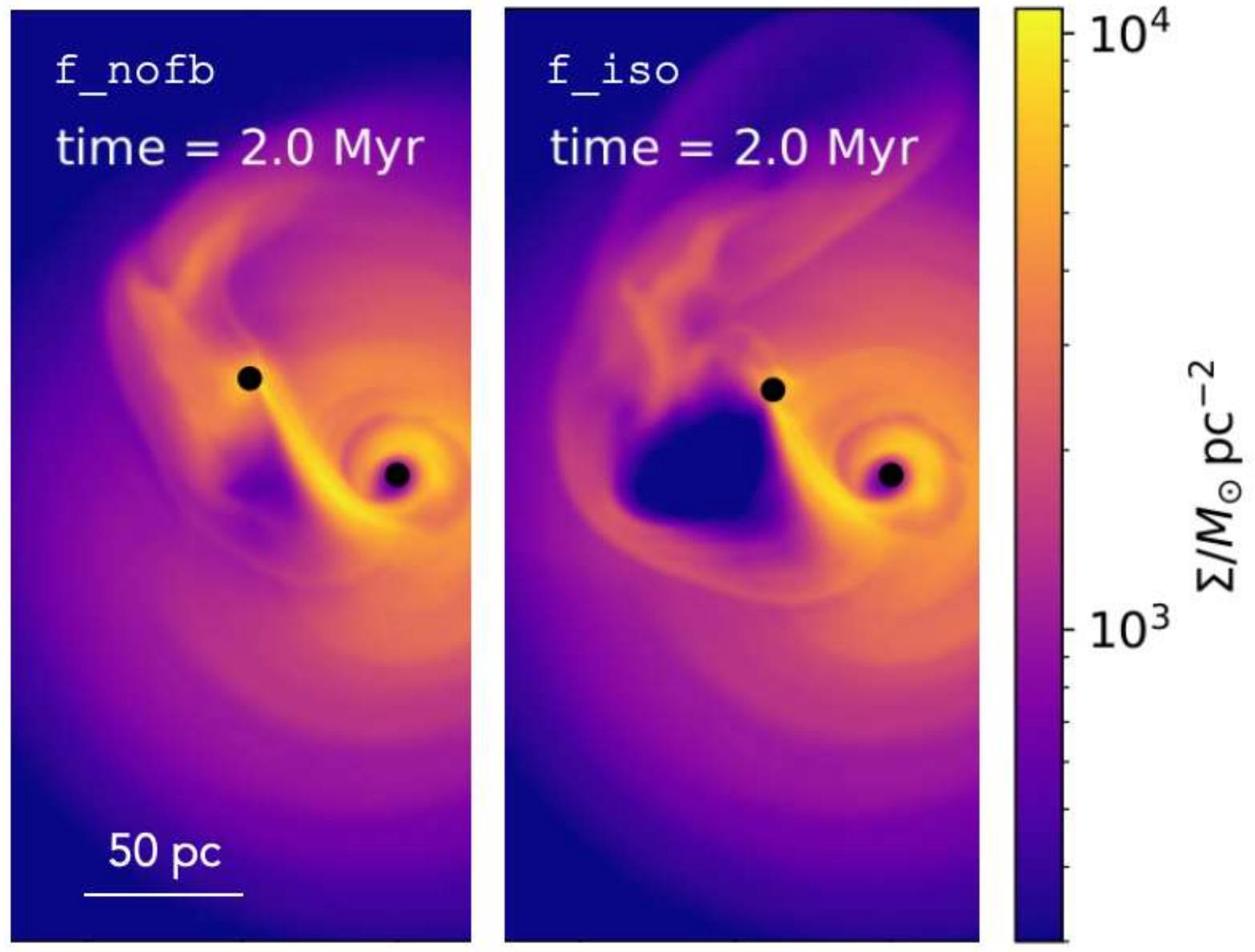}
    \caption{Two snapshots at time $t=2$ Myr of runs {\labsim{f\_nofb}} (left) and {\labsim{f\_iso}} (right).}
    \label{nofbVSiso}
\end{figure}

We can now use $A_2$ to interpret the dynamics of $M_2$ described in \Cref{fsim}. Fig.~\ref{Asf} shows the evolution of $A_2$ for all four {\labsim{f}}-simulations.
First, we observe that in {\labsim{nofb}} $A_2$ is positive, meaning that an overdensity is present behind the MBH. This produces a negative torque that forces $M_2$ to rapidly inspiral towards $M_1$. Conversely, when feedback is switched on, $A_2$ has initially positive values (and larger compared to {\labsim{nofb}}), suggesting that DF is enhanced in the early inspiral phase by the action of feedback (see. \Cref{fsim}). In Fig.~\ref{nofbVSiso} we compare snapshots taken at the same time (2~Myr after the start of the simualtion) for {\labsim{f\_nofb}} and {\labsim{f\_iso}}. In the first case, the formation of a spiral wave in the disc is accompanied by the presence of a moderately low-density region in front of $M_2$. When feedback is included, this region exerts a weaker resistance to 
the gas particles blown away by radiation pressure, allowing the bubble to expand in such direction. 
As a consequence, the gas surrounding $M_2$ exhibits a larger anisotropy, i.e., an initially larger value of $A_2$ corresponding to a larger negative torque. From  Fig.~\ref{Asf}, we also notice that in {\labsim{f\_iso}} the phase during which the bubble lies in front of $M_2$ ($A_2>0$) lasts longer compared to the other feedback models, consistently with the more prolonged negative torque observed in Fig.~\ref{torquesf}. Indeed, isotropic feedback is more efficient than anisotropic models ({\labsim{z}} and {\labsim{a}}) in keeping the low density bubble ``open'', as particles are more easily kicked in the CND plane (where the gas density is higher), hence more likely able to prolong the dynamical effect of the preceding bubble.

After the initial preceding-bubble ($A_2>0$) phase, all simulations with feedback show a  drop and eventually a sign change in $A_2$. 
This turning point approximately corresponds to $M_2$ reaching the pericenter. Indeed, as $M_2$ approaches the pericenter, its orbital speed exceeds that of the gas, overtaking the bubble which then lags behind. In this configuration, $A_2$ is negative, ans $M_2$ accelerates, increasing its eccentricity. In particular, {\labsim{f\_z}} displays the most negative value of $A_2$, implying that $M_2$ receives more energy in the process. In this case, $M_2$ reaches the first apocenter at a distance which is actually larger than the initial MBHs separation.
In {\labsim{f\_iso}} $A_2$ remains negative in the time interval $\simeq [3-10]$ Myr,, i.e., $M_2$ keeps being accelerated by the trailing bubble. Interestingly, in our simulations such feedback-driven acceleration is approximately balanced by stellar DF, and the semi-major axis remains approximately constant during this phase (see Fig.~\ref{A1fig} for a snapshot from this evolutionary phase). 

\subsection{{\labsim{e}}-simulations}\label{ce}

\begin{figure}
    \hspace{-0.043\textwidth}
    \includegraphics[scale=0.55]{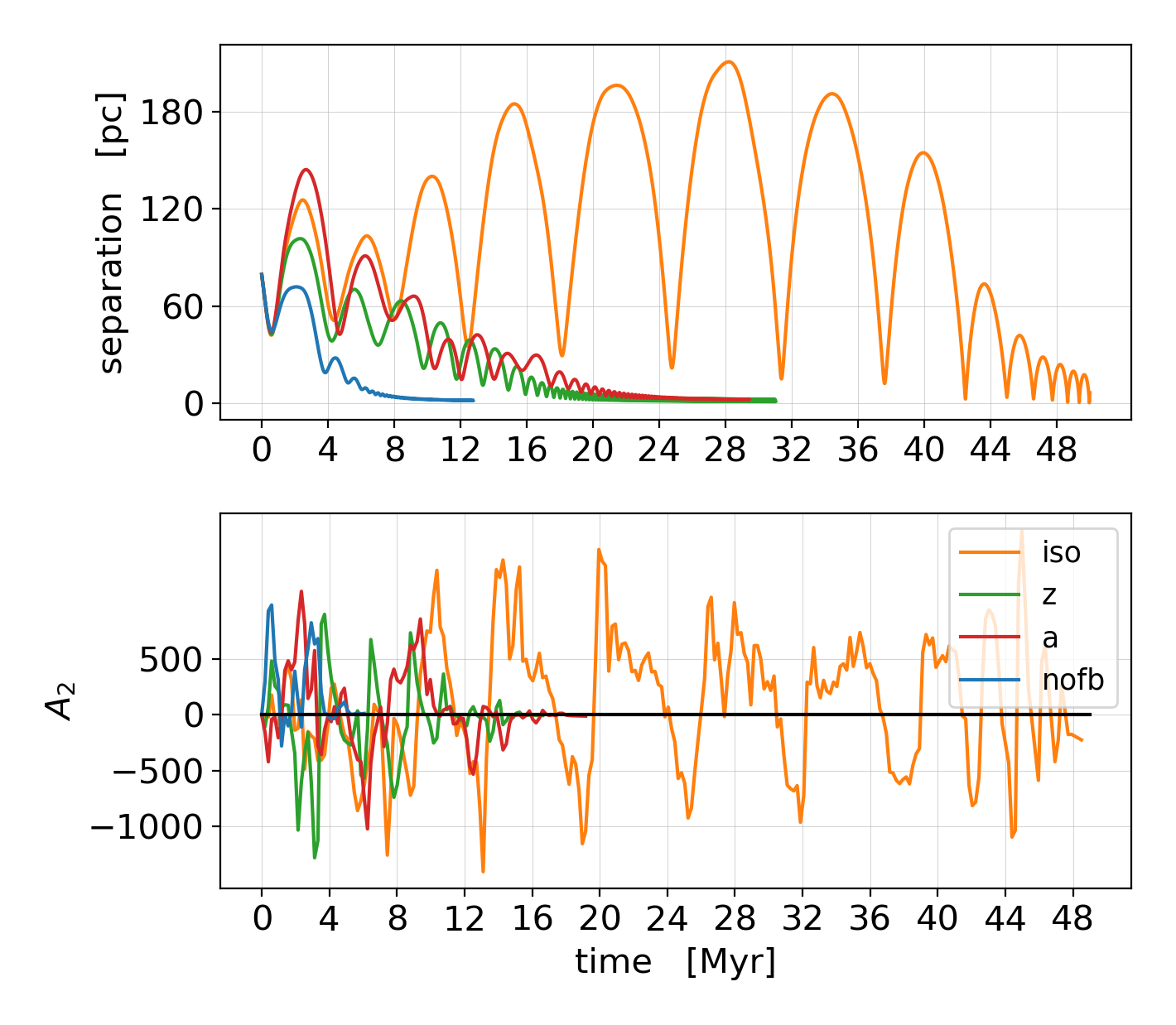}
    \caption{Time evolution of the MBHs separation (top) and anisotropy projection $A_2$ (bottom) in {\fontfamily{qcr}\selectfont\textbf{e}}-simulations.}
    \label{eDists}
\end{figure}

$M_2$ orbits corresponding to simulations with an eccentric initial condition are shown in Fig.~\ref{eDists} (top panel), together with the evolution of $A_2$ (bottom panel). In  {\labsim{e\_nofb}}, the MBH separation rapidly decays and orbit circularizes. This occurs as $M_2$ produces a density wake on the side opposite to the relative velocity between itself and the fluid. Therefore, since the orbital speed of $M_2$ close to the pericenter is larger than the local gas rotational velocity, the wake lags behind and $M_2$ slows down. On the other hand, near the apocenter the MBH velocity is smaller than the disc one and the wake is dragged in front of $M_2$, increasing its angular momentum and accelerating it. The combination of these two opposite effects at pericenter and apocenter results in orbit circularization \citep{Dotti2007,Bonetti2020}.
When radiative feedback is switched on, the density wake is somewhat destroyed and a low density bubble is created instead. Circularization is thus less effective, as it can be seen in  Fig.~\ref{eDists}, top panel, in the cases of {\labsim{e\_z}} and {\labsim{e\_a}}.
In {\labsim{e\_iso}}, because of the stronger impact of feedback on the surrounding gas, the density wake is more efficiently blown away and replaced by a low density bubble, which now follows the same trend of the wake in {\labsim{e\_nofb}}, but with the opposite gravitational effect. Therefore, at apocenter the bubble falls in front of the MBH, enhancing DF, while at pericenter it trails behind, accelerating the MBH, with the net effect of increasing the eccentricity. This behavior is illustrated in  Fig.~\ref{eDists} where, in {\labsim{e\_iso}}, $M_2$ develops relatively high ($ 0.5 \lesssim e \lesssim 0.9$) eccentricities. Correspondingly, $A_2$ is positive (i.e., bubble lies ahead) at apocenter  and negative (i.e., bubble lies behind) at pericenter, supporting our interpretation  (see Fig.\ref{eDists}, bottom panel).

\subsection{{\labsim{q}}-simulations}\label{cq}
\begin{figure}
    \hspace{-0.05\textwidth}
    \includegraphics[scale=0.55]{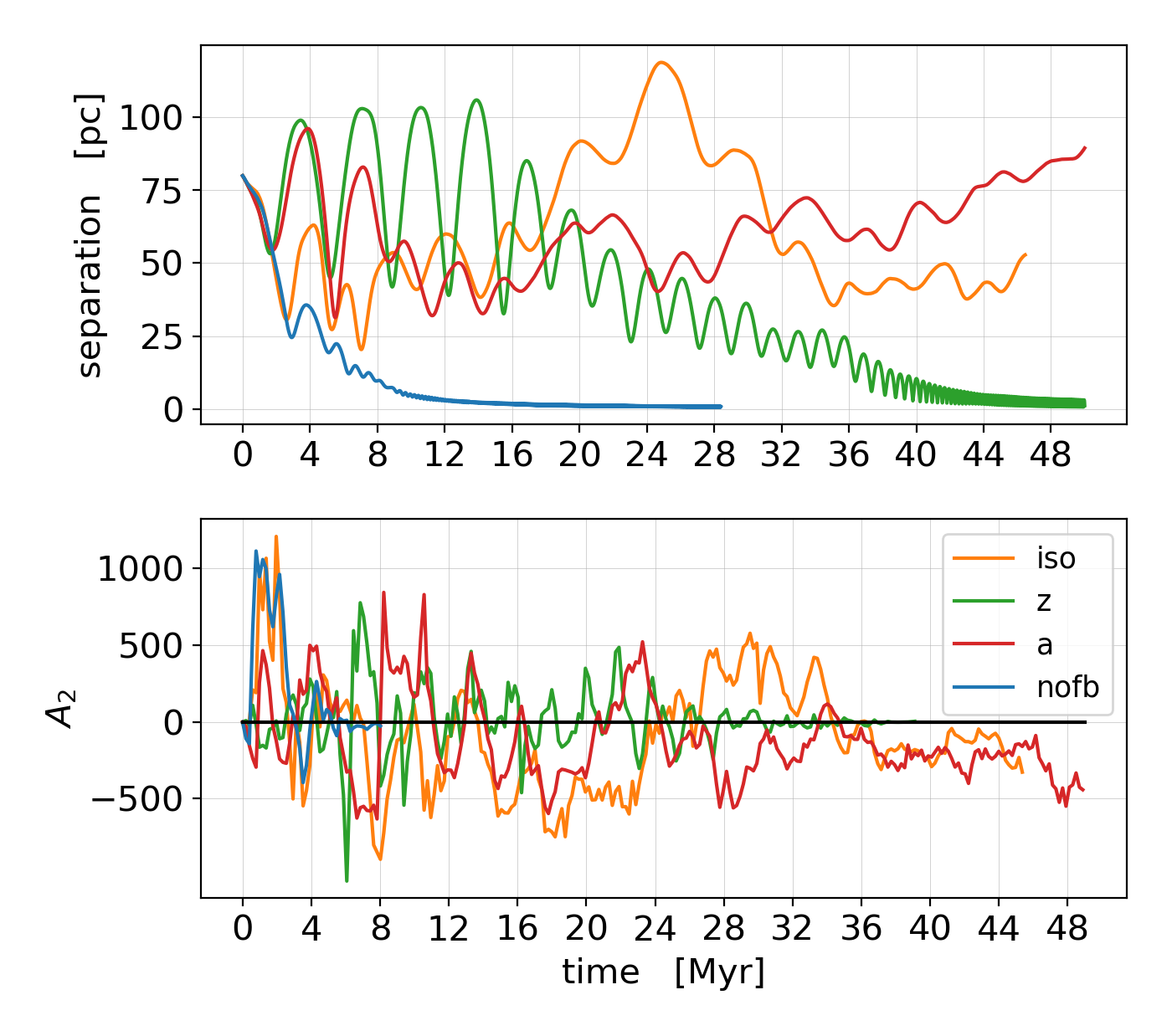}
    \caption{Same as Fig.~\ref{eDists} but for {\fontfamily{qcr}\selectfont\textbf{q}}-simulations.}
    \label{qDists}
\end{figure}

Fig.~\ref{qDists} illustrates the time evolution of MBH separation and of $A_2$ is illustrated in the case of a $q=1/6$ mass ratio. Overall, we can see that, in the feedback runs, $M_2$ orbits differ more from the {\labsim{nofb}} case, compared to what we have seen in {\labsim{f}} and {\labsim{e}} cases. The evolution of the orbital separation is again associated with the effects induced by the feedback. In {\labsim{q\_z}}, $M_2$ eccentricity increases up to $\sim 0.5$ in the first $\sim 15$ Myr, with the growth associated to an oscillating behaviour of $A_2$, positive at apocenter and negative at pericenter, as discussed in \cref{ce}. By contrast, in  {\labsim{q\_iso}} and {\labsim{q\_a}} $M_2$ orbits are quasi-circular, with an average eccentricity $e \lesssim 0.1$. In these cases, $M_2$ orbit is not going to shrink appreciably by the end of the simulation.  
Again, the increasing/decreasing trends of the MBHs separation due to positive/negative torques are linked to feedback, as they correspond, respectively, to negative/positive values of $A_2$ ( Fig.~\ref{qDists}, bottom).

Interestingly, in {\labsim{q\_a}}, after $\simeq 25$ Myr the distance of $M_2$ from $M_1$ shows, on average, a slightly increasing trend. We can explain this behavior as follows: due to the pressure gradient in the disc, the gas circular velocity is smaller than that of the MBH, settled on a quasi-circular orbit, and hence the bubble created in the disc by feedback is overtaken by the MBH, favoring its acceleration. In principle, this acceleration can be balanced by stellar DF, that concurs in  maintaining $M_2$ on a quasi-circular orbit. 


%
%


\subsection{Angular pattern}\label{ap}

In our analysis, we have shown through the quantity $A_2$ how radiative feedback is linked to the torques experienced by $M_2$, and how it can have an impact on orbital eccentricity and decay timescale. Here, we will discuss in a more quantitative way the relation existing between gas density perturbations due to feedback and the orbital decay timescale and eccentricity. 

If we consider the angle $\theta$ between $\Delta{\mathbf A}$ and $\mathbf{v}_2$, we do see that in the absence of feedback $M_2$ creates a trailing density wake, i.e., $\Delta{\mathbf A}$ tends to be directed parallel to $\mathbf{v}_2$, resulting in $\theta \lesssim \pi/2$. On the other hand, when feedback effects are considered, the low density bubble in the disc can be either trailing or leading with respect to $M_2$, depending on the relative velocity between the disc and $M_2$. This corresponds to $\theta\gtrsim \pi/2$ and $\theta\lesssim \pi/2$, respectively. 

\begin{figure}
    \hspace{0.02\textwidth}
    \includegraphics[scale=0.5]{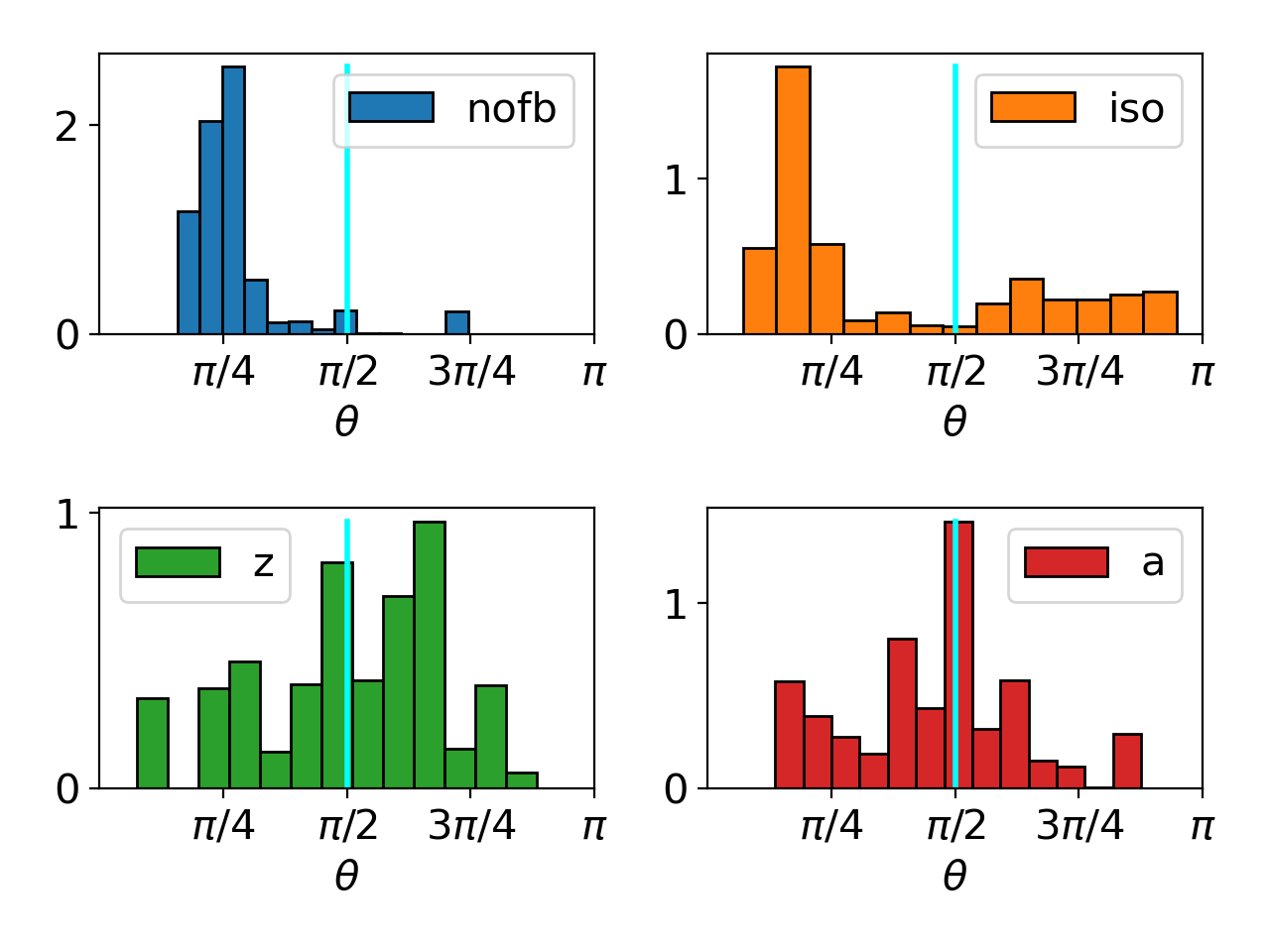}    \caption{Probability distribution of the (weighted) angle $\theta$ between the anisotropy difference $\mathbf{\Delta A}$ and $M_2$ velocity $\mathbf{v}_2$ (see text for details) in the  {\labsim{f}}-simulations.}
    \label{histoq}
\end{figure}

For all simulations, we follow the time evolution of $\theta$, 
and compute its weighted probability distribution. In practice, any occurrence of a given angle is weighted with the current value of $|\Delta \mathbf{A}|$, and the resulting frequency distribution of $\theta$ is then normalised. 
Results are shown in Fig.~\ref{histoq} 
for {\labsim{f}} simulations (the cases {\labsim{q}} and {\labsim{e}} are reported in \cref{apptheta}). As expected,  the distribution in {\labsim{nofb}} is peaked at small angles ($\lesssim \pi/2$), whereas the feedback cases exhibit much more spread values,  across the entire range. In particular, the more the peak of the distribution shifts to larger values, 
the more frequently the secondary will be accelerated by feedback, making DF inefficient. Therefore we consider $\langle \theta \rangle$, the mean of $\theta$, as a proxy for DF efficiency and, for each simulation, we compare its value with the orbital decay timescale and the mean eccentricity.

\begin{figure}
    \hspace{-0.02\textwidth}
    \includegraphics[scale=0.45]{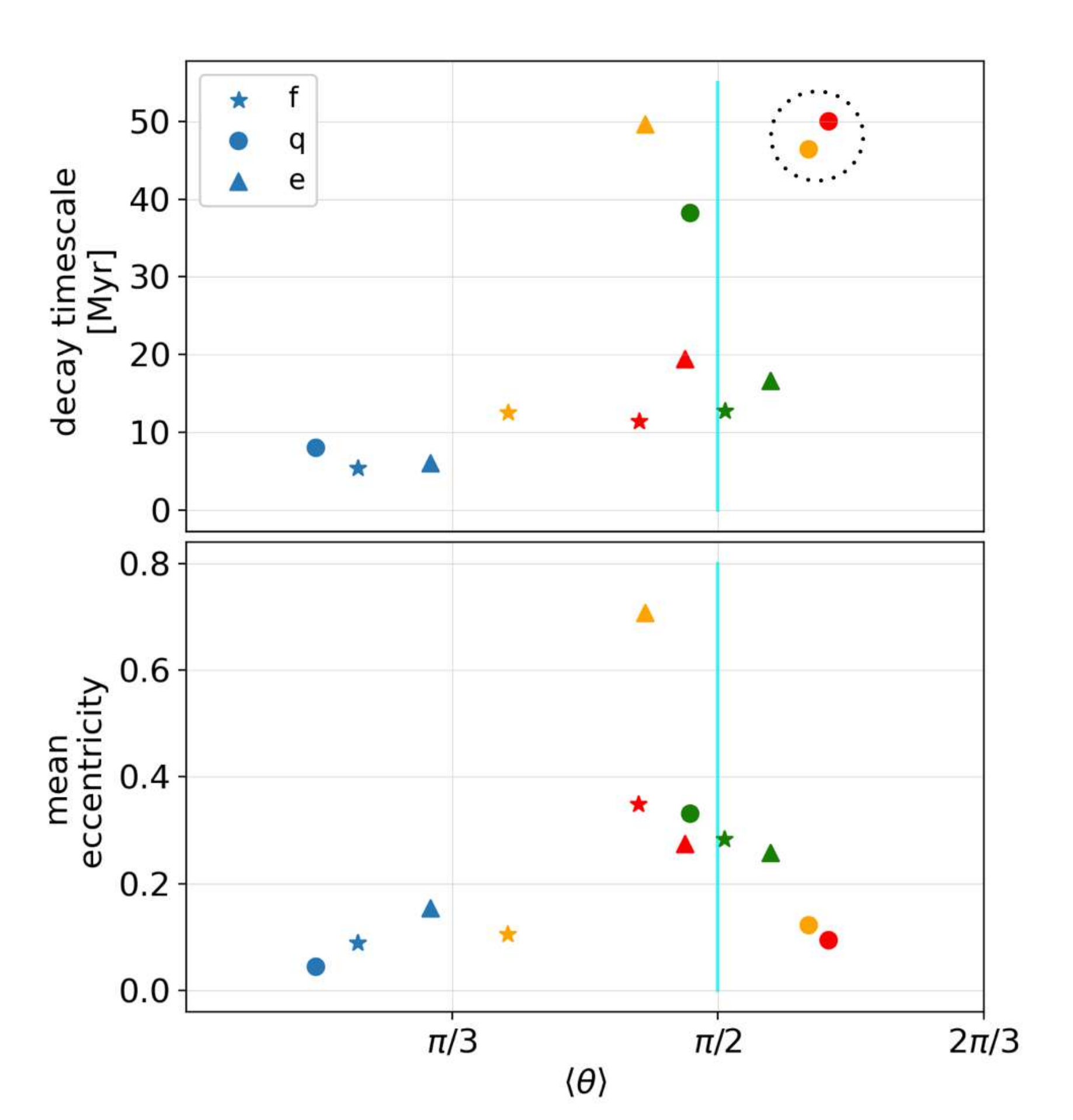}
    \caption{(\emph{top}) The decay timescale of $M_2$ vs the mean value of the angle $\theta$ between the anisotropy difference $\mathbf{\Delta A}$ and $M_2$ velocity $\mathbf{v}_2$, for the entire simulation suite. The decay time is estimated as the time that the semi-major axis of $M_2$ takes to reach an orbital distance to $M_1$ of 10 pc. 
    The different colors label the different type of feedback with the same legend of previous figures. Note that for the two runs in the dotted circle ({\labsim{q\_a}} and {\labsim{q\_z}}) the time indicated is the stop time of the simulation, as in these two cases $M_2$ did not reach an orbital separation of 10 pc within the simulation time.
    (\emph{bottom}) The mean eccentricity of $M_2$ vs the mean of $\theta$. The legend is the same as in top panel.
    }
    \label{AngleTime}
\end{figure}

In Fig.~\ref{AngleTime} (top panel), we plot $\langle \theta \rangle$ against the orbital decay timescale, defined here as the time required by $M_2$ to reach an orbital semi-major axis $<10$ pc. Simulations without feedback present lower mean values of $\theta$ and lower values of decaying timescale, while both quantities are larger in feedback simulations, confirming that a feedback-induced trailing bubble delays the inspiral of $M_2$ toward $M_1$.

In particular, if we compare {\labsim{f}} and {\labsim{q}} simulations, both with initial circular orbits but different mass ratios, we observe that by lowering the mass ratio by a factor of $1/3$ (i.e., moving from {\labsim{f}} to {\labsim{q}}) the decay timescale is significantly delayed. This indicates that feedback is more likely to affect lighter MBHs dynamics, or, in other words, that the feedback accelerating force has a softer scaling with the perturber mass $M$ compared to the DF force (which is $\propto M^2$). This is consistent with \cite{2020MNRAS.492.2755G}, who showed that the feedback-induced force acting on a pertuber moving in an homogeneous medium scales as $\propto M^{3/2}$, even though the underlying assumptions on the wind are different, as they assumed that the shocked wind thermal energy was instantaneously radiated away, whereas our simulations do not included radiative cooling at all.

Similarly, in Fig.~\ref{AngleTime} (bottom panel) we compare $\langle \theta \rangle$ with the mean eccentricity. In the absence of feedback, DF is efficient and both $\langle \theta \rangle$ and mean eccentricity are small ($< 0.2$), even for {\labsim{e\_nofb}}, which started eccentric. On the other hand, when feedback is turned on, the majority of simulations exhibits excited eccentricities (or hindered circularization, as in {\labsim{e\_z}} or {\labsim{e\_a}}). By contrast, in three cases ({\labsim{f\_iso}}, {\labsim{q\_iso}} and {\labsim{q\_z}}), the secondary mean eccentricity remains small ($\sim 0.1$), comparable with those found without feedback. The reason behind this different behaviour is that these simulations are characterized by prolonged stages of trailing bubbles in which feedback acceleration is counteracted by stellar DF.



\section{Summary and Conclusions}
\label{sec:summary}

Using the publicly available code \textsc{gizmo}, we have run a suite of simulations aimed at studying the dynamics of a MBH pair embedded in a gaseous circumnuclear disk. By means of dedicated sub-resolution prescriptions, we could model the dynamics in the presence of spin-dependent feedback, and compare results obtained with different feedback models, and to a benchmark case with no feedback at all. Overall, we observed that feedback significantly alters the MBHs dynamics and different feedback models produce large differences in the orbital evolution of the MBH pair.

Our results can be summarized as follows:

\begin{itemize}
\item feedback can both raise and suppress DF with the net effect of delaying the shrinking timescale of a MBH pair. This very fact bears important consequences when we are to model the cosmic population of MBH binaries, and their detectability via electromagnetic {\it and} gravitational waves;
\item feedback has also a relevant effect on the eccentricity of MBHs orbits, exciting it or weakening the circularization process.  
Again, this fact is relevant for an assessment of the properties of MBH pairs as a cosmological population; 
\item finally, the effect of feedback on the dynamics of MBH pairs is more pronounced for decreasing mass ratios. 

\end{itemize}


Our results then highlight that MBHs dynamics strongly depends on the feedback model assumed, i.e. whether we are considering isotropic or anisotropic feedback, with fixed or spin-dependent collimation axis. It is therefore crucial to model the anisotropy and direction of feedback consistently with the MBH spin, which, in turn, evolves according to the accretion on the MBH, in order to reliable assess the role of feedback in the evolution of MBH pairs. This is the only way to consistently capture the interplay between feeding and feedback, allowing a proper modeling of MBHs pairing, which is essential in view of forthcoming low-frequency GW missions such as LISA \citep{lisa}.


Due to our simplified modeling, a number of caveats that we have to keep in mind when interpreting our results do exist. Specifically:

\begin{itemize}
\item Our analysis is limited to coplanar orbits, an assumption justified by the fact that both the CBD and the MBHs inherit their angular momentum from the parent merging galaxies, leading to orbits likely laying in the same orbital plane \citep{Mayer07, Colpi07}. Nontheless, if coplanarity is not guaranteed, the pair inspiral is initially driven mainly by the DF exerted by the stellar bulge, while gaseous DF dominates once the MBHs separation becomes small enough that the MBHs spend most of their time in the disc \citep{escala05}. As a consequence, for non-coplanar orbits, we expect the feedback-induced eccentricity and delayed decay to be initially negligible and to become progressively more important as the pair shrinks and gets closer to binary formation.

\item In {\labsim{z}} and {\labsim{a}} feedback models
the anisotropy has been realized by kicking the gas particles within a well-defined cone, as if the driving radiation emitted by the subgrid disc had a step-like angular pattern, non-vanishing within the cone. 
In reality, the disc radiation angular pattern can be described with a continuous function which depends upon the MBH spin \citep{Campitiello18} and therefore evolves with it;

\item we have not included any form of cooling in the simulations. Cooling can potentially make the low density bubbles expand as momentum-driven structures, hence more slowly, since the shock wind thermal energy that swells the bubbles is radiated away. Therefore, cooling may have an impact on the bubbles formation/expansion which, in turn, may reflect on the MBH dynamics;

\item the MBH wind has been simulated via injection of kinetic energy only,
by adding momentum to gas particles within the MBH kernel. The lack of thermal energy injection tends to postpone the wind thermalization \citep{Costa20}, thus speeding up the bubble expansion, which, again, may affect the MBH dynamics;

\item If the region within $R_{\bullet,\textrm{max}}$ (the maximum MBH kernel size) is emptied, i.e. a sufficiently large low density bubble is formed around the MBH, feedback is artificially shut off since no more particles are eligible to be kicked, until the MBH kernel is refilled. This can weaken bubbles expansion thus influencing the MBH orbital evolution.

\end{itemize}

In a paper in preparation we are going to refine our recipes for feedback implementation, in order to overcome most of the the aforementioned limitations.
More in detail, our new implementation is based on spawning AGN wind gas particles \citep{Torrey20} from the subgrid accretion disc, such that their angular mass distribution follows the same angular pattern of the accretion disc luminosity \citep{Campitiello18}, in this way linking the wind anisotropy, and not only its direction, to the MBH spin. Then, wind particles are kicked outward radially at fixed velocity and by interacting with the surrounding gas particles they generate an anisotropic outflow. We expect this different feedback model to affect the formation of low density bubbles and hence the MBHs inspiral rates, compared to the present work. Indeed, on the one hand this new model would tend to produce wider bubbles because feedback is not artificially shut off once gas particles escape the BH kernel and wind launching covers the whole solid angle, irrespective of anisotorpy. On the other hand, the kinetic energy injection feedback model employed in this paper would be prone to generate stronger feedback, due to the delayed wind thermalisation intrinsic to this approach, than achieved via launching spawned wind particles. It is therefore not obvious how and by how much bubble formation and MBHs dynamics would differ due to using these different feedback subgrid models. We will address this issue in future works. 
\\


\section*{Acknowledgements}
We acknowledge the CINECA award under the ISCRA initiative, for the availability of high-performance computing resources and support (project number HP10CJ7AUZ).The analyses reported
in this work have been mainly performed using pynbody \citep{pynbody}.
A.L., M.D., \& F.H.  acknowledge funding from MIUR under the grant PRIN 2017-MB8AEZ.

\section*{Data Availability}
The data underlying this article will be shared on reasonable request
to the corresponding author.



\bibliographystyle{mnras}
\bibliography{biblio} 

\begin{thebibliography}{}
\makeatletter
\relax
\def\mn@urlcharsother{\let\do\@makeother \do\$\do\&\do\#\do\^\do\_\do\%\do\~}
\def\mn@doi{\begingroup\mn@urlcharsother \@ifnextchar [ {\mn@doi@}
  {\mn@doi@[]}}
\def\mn@doi@[#1]#2{\def\@tempa{#1}\ifx\@tempa\@empty \href
  {http://dx.doi.org/#2} {doi:#2}\else \href {http://dx.doi.org/#2} {#1}\fi
  \endgroup}
\def\mn@eprint#1#2{\mn@eprint@#1:#2::\@nil}
\def\mn@eprint@arXiv#1{\href {http://arxiv.org/abs/#1} {{\tt arXiv:#1}}}
\def\mn@eprint@dblp#1{\href {http://dblp.uni-trier.de/rec/bibtex/#1.xml}
  {dblp:#1}}
\def\mn@eprint@#1:#2:#3:#4\@nil{\def\@tempa {#1}\def\@tempb {#2}\def\@tempc
  {#3}\ifx \@tempc \@empty \let \@tempc \@tempb \let \@tempb \@tempa \fi \ifx
  \@tempb \@empty \def\@tempb {arXiv}\fi \@ifundefined
  {mn@eprint@\@tempb}{\@tempb:\@tempc}{\expandafter \expandafter \csname
  mn@eprint@\@tempb\endcsname \expandafter{\@tempc}}}

\bibitem[\protect\citeauthoryear{{Amaro-Seoane} et~al.,}{{Amaro-Seoane}
  et~al.}{2022}]{LISAW}
{Amaro-Seoane} P.,  et~al., 2022, arXiv e-prints, \href
  {https://ui.adsabs.harvard.edu/abs/2022arXiv220306016A} {p. arXiv:2203.06016}

\bibitem[\protect\citeauthoryear{{Angl{\'e}s-Alc{\'a}zar}, {Dav{\'e}},
  {Faucher-Gigu{\`e}re}, {{\"O}zel}  \& {Hopkins}}{{Angl{\'e}s-Alc{\'a}zar}
  et~al.}{2017}]{DAA2017}
{Angl{\'e}s-Alc{\'a}zar} D.,  {Dav{\'e}} R.,  {Faucher-Gigu{\`e}re} C.-A.,
  {{\"O}zel} F.,   {Hopkins} P.~F.,  2017, \mn@doi [\mnras]
  {10.1093/mnras/stw2565}, \href
  {https://ui.adsabs.harvard.edu/abs/2017MNRAS.464.2840A} {464, 2840}

\bibitem[\protect\citeauthoryear{{Bardeen} \& {Petterson}}{{Bardeen} \&
  {Petterson}}{1975}]{BP75}
{Bardeen} J.~M.,  {Petterson} J.~A.,  1975, \mn@doi [\apjl] {10.1086/181711},
  195, L65

\bibitem[\protect\citeauthoryear{{Barnes}}{{Barnes}}{2002}]{barnes02}
{Barnes} J.~E.,  2002, \mn@doi [\mnras] {10.1046/j.1365-8711.2002.05335.x},
  \href {https://ui.adsabs.harvard.edu/abs/2002MNRAS.333..481B} {333, 481}

\bibitem[\protect\citeauthoryear{{Barnes} \& {Hernquist}}{{Barnes} \&
  {Hernquist}}{1991}]{bh91}
{Barnes} J.~E.,  {Hernquist} L.~E.,  1991, \mn@doi [\apjl] {10.1086/185978},
  \href {https://ui.adsabs.harvard.edu/abs/1991ApJ...370L..65B} {370, L65}

\bibitem[\protect\citeauthoryear{{Barnes} \& {Hernquist}}{{Barnes} \&
  {Hernquist}}{1996}]{bh96}
{Barnes} J.~E.,  {Hernquist} L.,  1996, \mn@doi [\apj] {10.1086/177957}, \href
  {https://ui.adsabs.harvard.edu/abs/1996ApJ...471..115B} {471, 115}

\bibitem[\protect\citeauthoryear{{Begelman}, {Blandford}  \& {Rees}}{{Begelman}
  et~al.}{1980}]{BBT80}
{Begelman} M.~C.,  {Blandford} R.~D.,   {Rees} M.~J.,  1980, \mn@doi [\nat]
  {10.1038/287307a0}, 287, 307

\bibitem[\protect\citeauthoryear{{Blumenthal} \& {Barnes}}{{Blumenthal} \&
  {Barnes}}{2018}]{BB18}
{Blumenthal} K.~A.,  {Barnes} J.~E.,  2018, \mn@doi [\mnras]
  {10.1093/mnras/sty1605}, \href
  {https://ui.adsabs.harvard.edu/abs/2018MNRAS.479.3952B} {479, 3952}

\bibitem[\protect\citeauthoryear{{Bogdanovi{\'c}}, {Reynolds}  \&
  {Miller}}{{Bogdanovi{\'c}} et~al.}{2007}]{bogdanovic07}
{Bogdanovi{\'c}} T.,  {Reynolds} C.~S.,   {Miller} M.~C.,  2007, \mn@doi
  [\apjl] {10.1086/518769}, \href
  {https://ui.adsabs.harvard.edu/abs/2007ApJ...661L.147B} {661, L147}

\bibitem[\protect\citeauthoryear{{Bondi}}{{Bondi}}{1952}]{Bondi52}
{Bondi} H.,  1952, \mn@doi [\mnras] {10.1093/mnras/112.2.195}, 112, 195

\bibitem[\protect\citeauthoryear{{Bondi} \& {Hoyle}}{{Bondi} \&
  {Hoyle}}{1944}]{Bondi44}
{Bondi} H.,  {Hoyle} F.,  1944, \mn@doi [\mnras] {10.1093/mnras/104.5.273},
  104, 273

\bibitem[\protect\citeauthoryear{{Bonetti}, {Bortolas}, {Lupi}, {Dotti}  \&
  {Raimundo}}{{Bonetti} et~al.}{2020}]{Bonetti2020}
{Bonetti} M.,  {Bortolas} E.,  {Lupi} A.,  {Dotti} M.,   {Raimundo} S.~I.,
  2020, \mn@doi [\mnras] {10.1093/mnras/staa964}, \href
  {https://ui.adsabs.harvard.edu/abs/2020MNRAS.494.3053B} {494, 3053}

\bibitem[\protect\citeauthoryear{{Campitiello}, {Ghisellini}, {Sbarrato}  \&
  {Calderone}}{{Campitiello} et~al.}{2018}]{Campitiello18}
{Campitiello} S.,  {Ghisellini} G.,  {Sbarrato} T.,   {Calderone} G.,  2018,
  \mn@doi [\aap] {10.1051/0004-6361/201731897}, \href
  {https://ui.adsabs.harvard.edu/abs/2018A&A...612A..59C} {612, A59}

\bibitem[\protect\citeauthoryear{{Capelo} \& {Dotti}}{{Capelo} \&
  {Dotti}}{2017}]{CD17}
{Capelo} P.~R.,  {Dotti} M.,  2017, \mn@doi [\mnras] {10.1093/mnras/stw2872},
  \href {https://ui.adsabs.harvard.edu/abs/2017MNRAS.465.2643C} {465, 2643}

\bibitem[\protect\citeauthoryear{{Cenci}, {Sala}, {Lupi}, {Capelo}  \&
  {Dotti}}{{Cenci} et~al.}{2021}]{Cenci2021}
{Cenci} E.,  {Sala} L.,  {Lupi} A.,  {Capelo} P.~R.,   {Dotti} M.,  2021,
  \mn@doi [\mnras] {10.1093/mnras/staa3449}, 500, 3719

\bibitem[\protect\citeauthoryear{{Chandrasekhar}}{{Chandrasekhar}}{1943}]{Chandra43}
{Chandrasekhar} S.,  1943, \mn@doi [\apj] {10.1086/144517}, 97, 255

\bibitem[\protect\citeauthoryear{{Colpi}, {Dotti}, {Mayer}  \&
  {Kazantzidis}}{{Colpi} et~al.}{2007}]{Colpi07}
{Colpi} M.,  {Dotti} M.,  {Mayer} L.,   {Kazantzidis} S.,  2007, arXiv
  e-prints, \href {https://ui.adsabs.harvard.edu/abs/2007arXiv0710.5207C} {p.
  arXiv:0710.5207}

\bibitem[\protect\citeauthoryear{{Costa}, {Pakmor}  \& {Springel}}{{Costa}
  et~al.}{2020}]{Costa20}
{Costa} T.,  {Pakmor} R.,   {Springel} V.,  2020, \mn@doi [\mnras]
  {10.1093/mnras/staa2321}, \href
  {https://ui.adsabs.harvard.edu/abs/2020MNRAS.497.5229C} {497, 5229}

\bibitem[\protect\citeauthoryear{{Curtis} \& {Sijacki}}{{Curtis} \&
  {Sijacki}}{2016}]{Curtis16}
{Curtis} M.,  {Sijacki} D.,  2016, \mn@doi [\mnras] {10.1093/mnras/stw1944},
  \href {https://ui.adsabs.harvard.edu/abs/2016MNRAS.463...63C} {463, 63}

\bibitem[\protect\citeauthoryear{{De Rosa} et~al.,}{{De Rosa}
  et~al.}{2019}]{derosa}
{De Rosa} A.,  et~al., 2019, \mn@doi [\nar] {10.1016/j.newar.2020.101525},
  \href {https://ui.adsabs.harvard.edu/abs/2019NewAR..8601525D} {86, 101525}

\bibitem[\protect\citeauthoryear{{Dotti}, {Colpi}  \& {Haardt}}{{Dotti}
  et~al.}{2006}]{dotti06}
{Dotti} M.,  {Colpi} M.,   {Haardt} F.,  2006, \mn@doi [\mnras]
  {10.1111/j.1365-2966.2005.09956.x}, \href
  {https://ui.adsabs.harvard.edu/abs/2006MNRAS.367..103D} {367, 103}

\bibitem[\protect\citeauthoryear{{Dotti}, {Colpi}, {Haardt}  \&
  {Mayer}}{{Dotti} et~al.}{2007}]{Dotti2007}
{Dotti} M.,  {Colpi} M.,  {Haardt} F.,   {Mayer} L.,  2007, \mn@doi [\mnras]
  {10.1111/j.1365-2966.2007.12010.x}, 379, 956

\bibitem[\protect\citeauthoryear{{Dotti}, {Ruszkowski}, {Paredi}, {Colpi},
  {Volonteri}  \& {Haardt}}{{Dotti} et~al.}{2009}]{dotti09}
{Dotti} M.,  {Ruszkowski} M.,  {Paredi} L.,  {Colpi} M.,  {Volonteri} M.,
  {Haardt} F.,  2009, \mn@doi [\mnras] {10.1111/j.1365-2966.2009.14840.x},
  \href {https://ui.adsabs.harvard.edu/abs/2009MNRAS.396.1640D} {396, 1640}

\bibitem[\protect\citeauthoryear{{Dotti}, {Volonteri}, {Perego}, {Colpi},
  {Ruszkowski}  \& {Haardt}}{{Dotti} et~al.}{2010}]{dotti10}
{Dotti} M.,  {Volonteri} M.,  {Perego} A.,  {Colpi} M.,  {Ruszkowski} M.,
  {Haardt} F.,  2010, \mn@doi [\mnras] {10.1111/j.1365-2966.2009.15922.x},
  \href {https://ui.adsabs.harvard.edu/abs/2010MNRAS.402..682D} {402, 682}

\bibitem[\protect\citeauthoryear{{Downes} \& {Solomon}}{{Downes} \&
  {Solomon}}{1998}]{DS98}
{Downes} D.,  {Solomon} P.~M.,  1998, \mn@doi [\apj] {10.1086/306339}, \href
  {https://ui.adsabs.harvard.edu/abs/1998ApJ...507..615D} {507, 615}

\bibitem[\protect\citeauthoryear{{Escala}, {Larson}, {Coppi}  \&
  {Mardones}}{{Escala} et~al.}{2005}]{escala05}
{Escala} A.,  {Larson} R.~B.,  {Coppi} P.~S.,   {Mardones} D.,  2005, \mn@doi
  [\apj] {10.1086/431747}, \href
  {https://ui.adsabs.harvard.edu/abs/2005ApJ...630..152E} {630, 152}

\bibitem[\protect\citeauthoryear{{Fiacconi}, {Sijacki}  \&
  {Pringle}}{{Fiacconi} et~al.}{2018}]{Fiacconi2018}
{Fiacconi} D.,  {Sijacki} D.,   {Pringle} J.~E.,  2018, \mn@doi [\mnras]
  {10.1093/mnras/sty893}, \href
  {https://ui.adsabs.harvard.edu/abs/2018MNRAS.477.3807F} {477, 3807}

\bibitem[\protect\citeauthoryear{{Gruzinov}, {Levin}  \& {Matzner}}{{Gruzinov}
  et~al.}{2020}]{2020MNRAS.492.2755G}
{Gruzinov} A.,  {Levin} Y.,   {Matzner} C.~D.,  2020, \mn@doi [\mnras]
  {10.1093/mnras/staa013}, \href
  {https://ui.adsabs.harvard.edu/abs/2020MNRAS.492.2755G} {492, 2755}

\bibitem[\protect\citeauthoryear{{Hernquist}}{{Hernquist}}{1989}]{hernquist89}
{Hernquist} L.,  1989, \mn@doi [\nat] {10.1038/340687a0}, \href
  {https://ui.adsabs.harvard.edu/abs/1989Natur.340..687H} {340, 687}

\bibitem[\protect\citeauthoryear{{Hernquist}}{{Hernquist}}{1990}]{H90}
{Hernquist} L.,  1990, \mn@doi [\apj] {10.1086/168845}, \href
  {https://ui.adsabs.harvard.edu/abs/1990ApJ...356..359H} {356, 359}

\bibitem[\protect\citeauthoryear{{Hopkins}}{{Hopkins}}{2015}]{Hopkins2015}
{Hopkins} P.~F.,  2015, \mn@doi [\mnras] {10.1093/mnras/stv195}, 450, 53

\bibitem[\protect\citeauthoryear{{Hopkins} \& {Quataert}}{{Hopkins} \&
  {Quataert}}{2011}]{Hopkins11}
{Hopkins} P.~F.,  {Quataert} E.,  2011, \mn@doi [\mnras]
  {10.1111/j.1365-2966.2011.18542.x}, \href
  {https://ui.adsabs.harvard.edu/abs/2011MNRAS.415.1027H} {415, 1027}

\bibitem[\protect\citeauthoryear{Hoyle \& Lyttleton}{Hoyle \&
  Lyttleton}{1939}]{HL39}
Hoyle F.,  Lyttleton R.~A.,  1939, \mn@doi [Mathematical Proceedings of the
  Cambridge Philosophical Society] {10.1017/S0305004100021150}, 35, 405–415

\bibitem[\protect\citeauthoryear{{King}, {Lubow}, {Ogilvie}  \&
  {Pringle}}{{King} et~al.}{2005}]{king2005}
{King} A.~R.,  {Lubow} S.~H.,  {Ogilvie} G.~I.,   {Pringle} J.~E.,  2005,
  \mn@doi [\mnras] {10.1111/j.1365-2966.2005.09378.x}, 363, 49

\bibitem[\protect\citeauthoryear{{Lodato} \& {Pringle}}{{Lodato} \&
  {Pringle}}{2007}]{LP2006}
{Lodato} G.,  {Pringle} J.~E.,  2007, \mn@doi [\mnras]
  {10.1111/j.1365-2966.2007.12332.x}, 381, 1287

\bibitem[\protect\citeauthoryear{{Lupi}, {Haardt}  \& {Dotti}}{{Lupi}
  et~al.}{2015}]{Lupi2015}
{Lupi} A.,  {Haardt} F.,   {Dotti} M.,  2015, \mn@doi [\mnras]
  {10.1093/mnras/stu2223}, 446, 1765

\bibitem[\protect\citeauthoryear{{Masset}}{{Masset}}{2017}]{masset17b}
{Masset} F.~S.,  2017, \mn@doi [\mnras] {10.1093/mnras/stx2271}, \href
  {https://ui.adsabs.harvard.edu/abs/2017MNRAS.472.4204M} {472, 4204}

\bibitem[\protect\citeauthoryear{{Masset} \& {Velasco Romero}}{{Masset} \&
  {Velasco Romero}}{2017}]{masset17}
{Masset} F.~S.,  {Velasco Romero} D.~A.,  2017, \mn@doi [\mnras]
  {10.1093/mnras/stw3008}, \href
  {https://ui.adsabs.harvard.edu/abs/2017MNRAS.465.3175M} {465, 3175}

\bibitem[\protect\citeauthoryear{{Mayer}, {Kazantzidis}, {Madau}, {Colpi},
  {Quinn}  \& {Wadsley}}{{Mayer} et~al.}{2007}]{Mayer07}
{Mayer} L.,  {Kazantzidis} S.,  {Madau} P.,  {Colpi} M.,  {Quinn} T.,
  {Wadsley} J.,  2007, \mn@doi [Science] {10.1126/science.1141858}, \href
  {https://ui.adsabs.harvard.edu/abs/2007Sci...316.1874M} {316, 1874}

\bibitem[\protect\citeauthoryear{{Mihos} \& {Hernquist}}{{Mihos} \&
  {Hernquist}}{1996}]{mh96}
{Mihos} J.~C.,  {Hernquist} L.,  1996, \mn@doi [\apj] {10.1086/177353}, \href
  {https://ui.adsabs.harvard.edu/abs/1996ApJ...464..641M} {464, 641}

\bibitem[\protect\citeauthoryear{{Ostriker}}{{Ostriker}}{1999}]{Ostriker99}
{Ostriker} E.~C.,  1999, \mn@doi [\apj] {10.1086/306858}, 513, 252

\bibitem[\protect\citeauthoryear{{Pontzen}, {Ro{\v{s}}kar}, {Stinson}  \&
  {Woods}}{{Pontzen} et~al.}{2013}]{pynbody}
{Pontzen} A.,  {Ro{\v{s}}kar} R.,  {Stinson} G.,   {Woods} R.,  2013, {pynbody:
  N-Body/SPH analysis for python}, Astrophysics Source Code Library, record
  ascl:1305.002 (\mn@eprint {ascl} {1305.002})

\bibitem[\protect\citeauthoryear{{Sala}, {Cenci}, {Capelo}, {Lupi}  \&
  {Dotti}}{{Sala} et~al.}{2021}]{Sala2021}
{Sala} L.,  {Cenci} E.,  {Capelo} P.~R.,  {Lupi} A.,   {Dotti} M.,  2021,
  \mn@doi [\mnras] {10.1093/mnras/staa3552}, 500, 4788

\bibitem[\protect\citeauthoryear{{Sanders} \& {Mirabel}}{{Sanders} \&
  {Mirabel}}{1996}]{SM96}
{Sanders} D.~B.,  {Mirabel} I.~F.,  1996, \mn@doi [\araa]
  {10.1146/annurev.astro.34.1.749}, \href
  {https://ui.adsabs.harvard.edu/abs/1996ARA&A..34..749S} {34, 749}

\bibitem[\protect\citeauthoryear{{Shakura} \& {Sunyaev}}{{Shakura} \&
  {Sunyaev}}{1973}]{SS73}
{Shakura} N.~I.,  {Sunyaev} R.~A.,  1973, \aap, \href
  {https://ui.adsabs.harvard.edu/abs/1973A&A....24..337S} {500, 33}

\bibitem[\protect\citeauthoryear{{Sijacki}, {Springel}  \&
  {Haehnelt}}{{Sijacki} et~al.}{2011}]{sijacki11}
{Sijacki} D.,  {Springel} V.,   {Haehnelt} M.~G.,  2011, \mn@doi [\mnras]
  {10.1111/j.1365-2966.2011.18666.x}, \href
  {https://ui.adsabs.harvard.edu/abs/2011MNRAS.414.3656S} {414, 3656}

\bibitem[\protect\citeauthoryear{{Souza Lima}, {Mayer}, {Capelo}  \&
  {Bellovary}}{{Souza Lima} et~al.}{2017}]{rafael17}
{Souza Lima} R.,  {Mayer} L.,  {Capelo} P.~R.,   {Bellovary} J.~M.,  2017,
  \mn@doi [\apj] {10.3847/1538-4357/aa5d19}, \href
  {https://ui.adsabs.harvard.edu/abs/2017ApJ...838...13S} {838, 13}

\bibitem[\protect\citeauthoryear{{Springel}, {Di Matteo}  \&
  {Hernquist}}{{Springel} et~al.}{2005}]{Springeletal2005}
{Springel} V.,  {Di Matteo} T.,   {Hernquist} L.,  2005, \mn@doi [\mnras]
  {10.1111/j.1365-2966.2005.09238.x}, 361, 776

\bibitem[\protect\citeauthoryear{{Torrey} et~al.,}{{Torrey}
  et~al.}{2020}]{Torrey20}
{Torrey} P.,  et~al., 2020, \mn@doi [\mnras] {10.1093/mnras/staa2222}, \href
  {https://ui.adsabs.harvard.edu/abs/2020MNRAS.497.5292T} {497, 5292}

\bibitem[\protect\citeauthoryear{{Toyouchi}, {Hosokawa}, {Sugimura}  \&
  {Kuiper}}{{Toyouchi} et~al.}{2020}]{2020MNRAS.496.1909T}
{Toyouchi} D.,  {Hosokawa} T.,  {Sugimura} K.,   {Kuiper} R.,  2020, \mn@doi
  [\mnras] {10.1093/mnras/staa1338}, \href
  {https://ui.adsabs.harvard.edu/abs/2020MNRAS.496.1909T} {496, 1909}

\bibitem[\protect\citeauthoryear{{Tremmel}, {Karcher}, {Governato},
  {Volonteri}, {Quinn}, {Pontzen}, {Anderson}  \& {Bellovary}}{{Tremmel}
  et~al.}{2017}]{Tremmel17}
{Tremmel} M.,  {Karcher} M.,  {Governato} F.,  {Volonteri} M.,  {Quinn} T.~R.,
  {Pontzen} A.,  {Anderson} L.,   {Bellovary} J.,  2017, \mn@doi [\mnras]
  {10.1093/mnras/stx1160}, \href
  {https://ui.adsabs.harvard.edu/abs/2017MNRAS.470.1121T} {470, 1121}

\bibitem[\protect\citeauthoryear{{Volonteri}, {G{\"u}ltekin}  \&
  {Dotti}}{{Volonteri} et~al.}{2010}]{volonteri10}
{Volonteri} M.,  {G{\"u}ltekin} K.,   {Dotti} M.,  2010, \mn@doi [\mnras]
  {10.1111/j.1365-2966.2010.16431.x}, \href
  {https://ui.adsabs.harvard.edu/abs/2010MNRAS.404.2143V} {404, 2143}

\bibitem[\protect\citeauthoryear{{eLISA Consortium} et~al.,}{{eLISA Consortium}
  et~al.}{2013}]{lisa}
{eLISA Consortium} et~al., 2013, arXiv e-prints, \href
  {https://ui.adsabs.harvard.edu/abs/2013arXiv1305.5720E} {p. arXiv:1305.5720}

\makeatother
\end{thebibliography}




\appendix

\section{Convergence}\label{Conv}
In order to discuss the convergence of our results we performed additional simulations, with both smaller and higher resolution, in the cases {\labsim{f\_nofb}} and {\labsim{f\_a}}. In particular, we sampled the CND with $10^5$, $3\cdot 10^5$ and $3\cdot10^6$ gas particles, in addition to the $10^6$ case presented in \cref{fsim}. 
For all these resolutions the number of star particles is five times that of the gas. 
\begin{figure}
    \centering
    \includegraphics[scale=0.48]{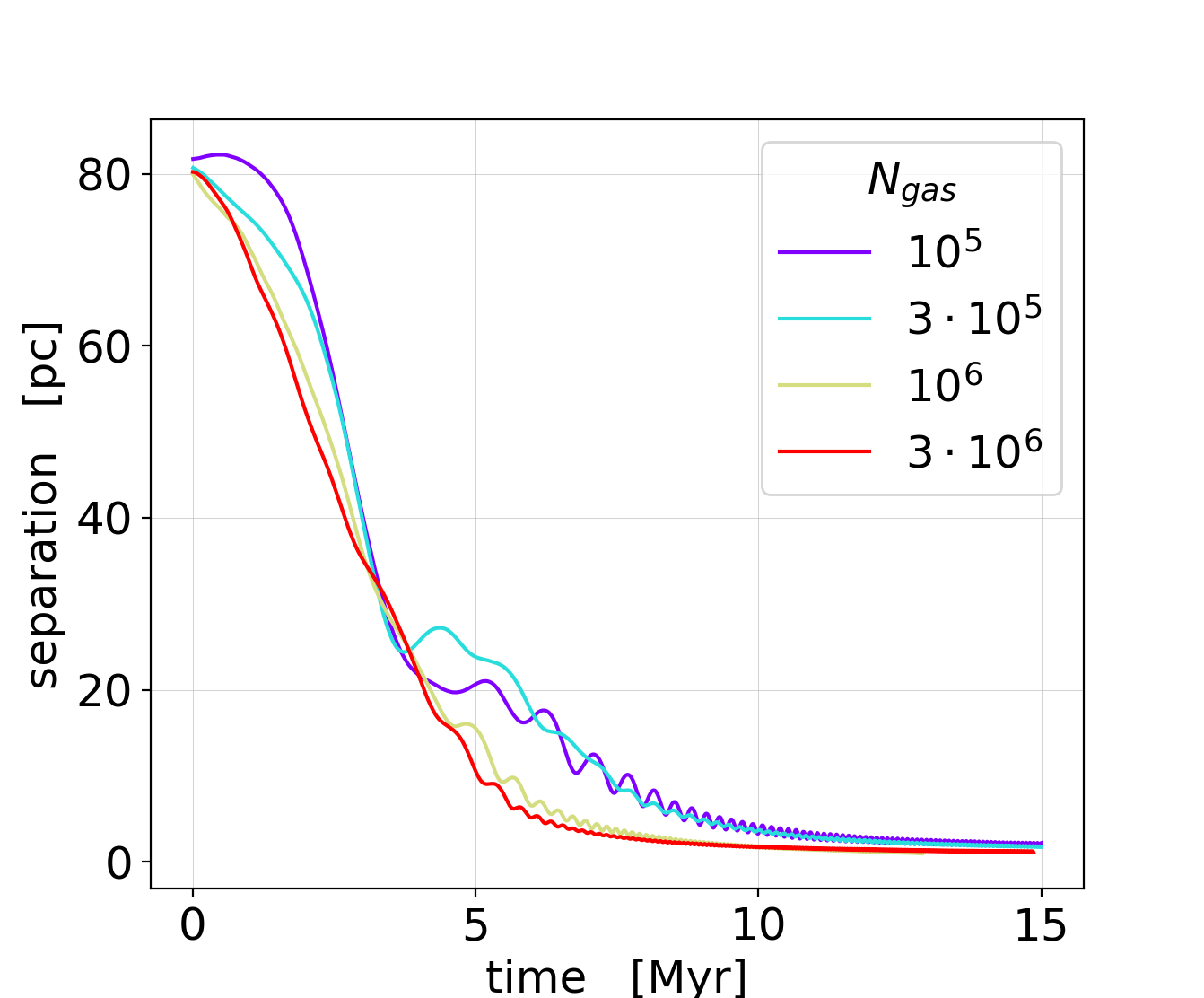}
    \caption{Time evoultion of the MBHs separation for different resolutions in the case {\labsim{f\_nofb}}.}
    \label{fig:resolution}
\end{figure}
\begin{figure}
    \hspace{-0.035\textwidth}
    \centering
    \includegraphics[scale=0.45]{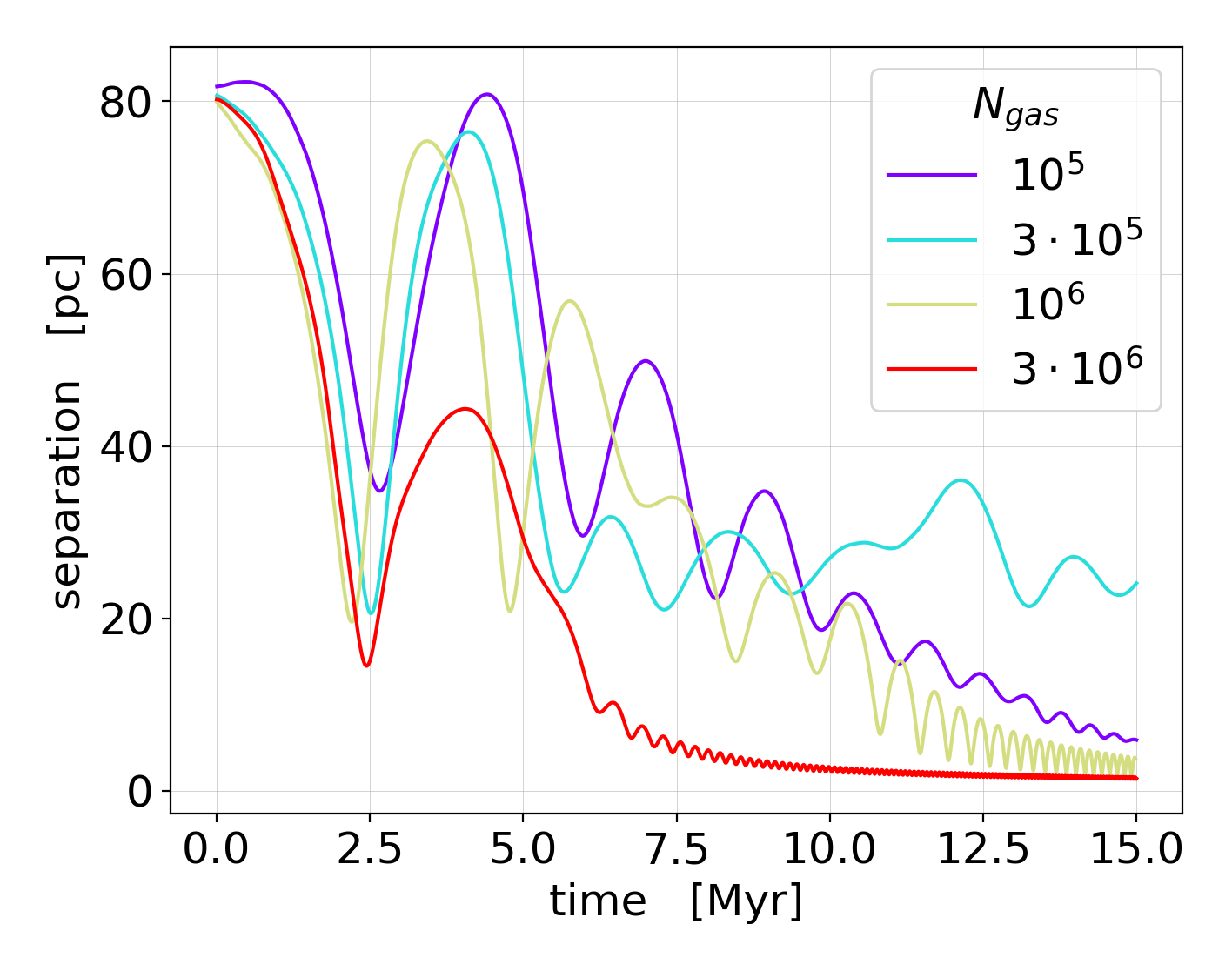}
    \caption{Time evoultion of the MBHs separation for different resolutions in the case {\labsim{f\_a}}.}
    \label{fig:resolution_a}
\end{figure}
Figure \ref{fig:resolution} shows how the evolution of the MBHs separation in {\labsim{f\_nofb}} changes with resolution and reveals that the dynamics converges by increasing the number of particles. On the other hand, we do not expect convergence in simulations with feedback due to the stochastic nature of our subgrid feedback model. Indeed, depending on which gas particles receive a kick at a given timestep, the resulting bubbles can display morphological differences, such that their cumulative effect in time can lead to very different trajectories, i.e. different realizations of the same stochastic process (see Fig \ref{fig:resolution_a} for the {\labsim{f\_a}} case).

\section{Force softening}\label{appw}
The weighting function $w$ appearing in the definition of the anisotropy vector (Eq. \ref{A11}) is defined as the force softening function used in \textsc{gizmo} (see Eq. (H6) in \cite{Hopkins2015}), that is $w(r) \equiv h^{-1}d\phi(q;h)/dq$ where $h = 2.8 \times \epsilon _\textrm{soft,BH}$ is the force softening length, $q=r/h$ and 
\begin{equation}
     \phi(q;h) = -\frac{1}{qh}\begin{cases}
             \frac{14}{5}q -\frac{16}{3}q^3 + \frac{48}{5}q^5 -\frac{32}{5}q^6, & 0\leq q < 0.5 \\
            -\frac{1}{15}+\frac{16}{5}q - \frac{32}{3}q^3 + 16q^4 - \frac{48}{5}q^5 + \frac{32}{15}q^6, & 0.5 \leq q < 1 \\
            1,        & q \geq 1.
          \end{cases}
\end{equation}

\section{Angular distributions}\label{apptheta}
\begin{figure}
    \centering
    \includegraphics[scale=0.5]{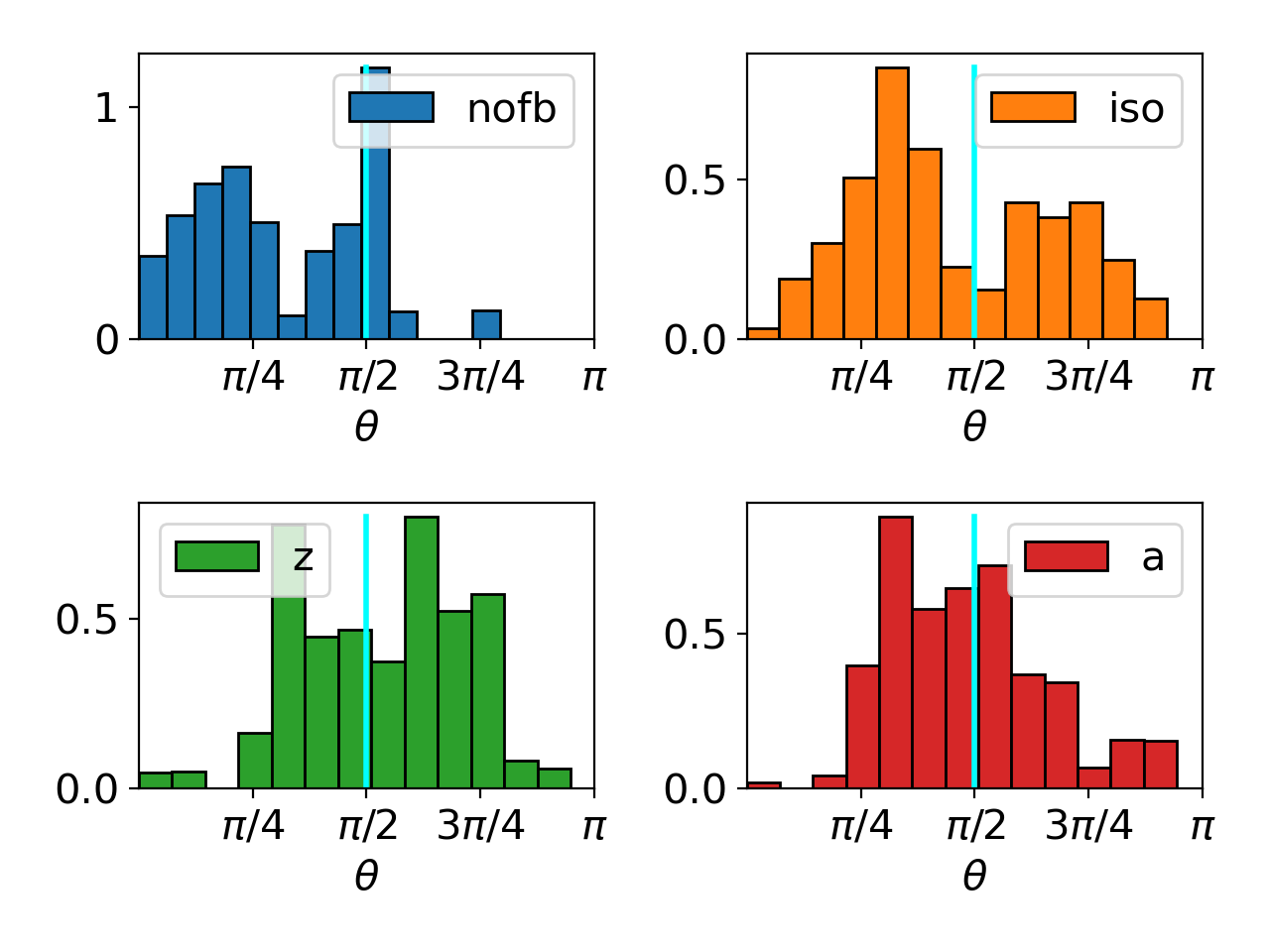}
    \caption{Histograms of $\theta$ in {\labsim{e}} simulations.}
    \label{ehisto}
\end{figure}

\begin{figure}
    \centering
    \includegraphics[scale=0.5]{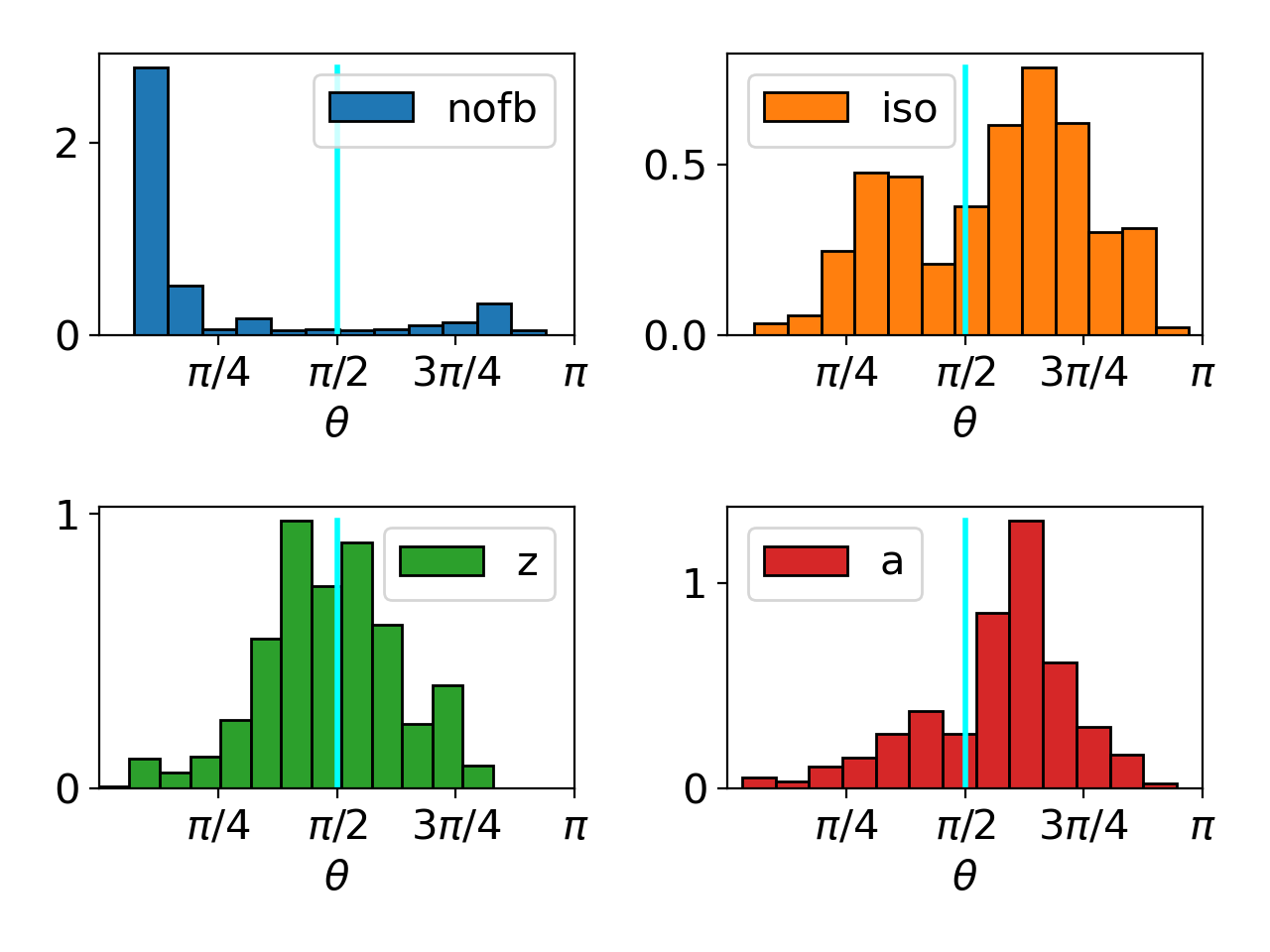}
    \caption{Histograms of $\theta$ in {\labsim{q}} simulations.}
    \label{qhisto}
\end{figure}

In Figs. \ref{qhisto} and \ref{ehisto} we report the histograms of the quantity $\theta$ for {\labsim{e}} and {\labsim{q}} simulations. As pointed out in \cref{ap}, the distribution of $\theta$, peaked at small angles ($<\pi/2$) in {\labsim{nofb}} simulations, spreads over the whole range $[0,\pi]$ when feedback is turned on, due to the presence of low density bubbles trailing the MBH, which tend to accelerate it, hampering DF. We note that the simulations in which the peak of the distribution is more shifted to the right (i.e. more frequent feedback acceleration) are {\labsim{q\_iso}} and {\labsim{q\_a}}, which are also the only two simulations whose orbits do not decay over the simulated time. 



\bsp	
\label{lastpage}
\end{document}